\documentclass[10pt,floatfix,letterpaper,amsmath,amsfonts,amssymb,aps,pra,reprint,longbibliography,superscriptaddress]{revtex4-1}

\usepackage{amsmath}
\usepackage{graphicx}
\usepackage[pdfpagemode=UseNone,pdfstartview=FitH,colorlinks=true,linkcolor=blue,citecolor=blue,urlcolor=blue]{hyperref}
\usepackage[all]{hypcap}

\newcommand{\ket}[1]{\lvert #1\rangle}					
\newcommand{\bra}[1]{\langle #1\rvert}					
\newcommand{\mean}[1]{\langle #1 \rangle}				

\def\be{\begin{equation}}
\def\ee{\end{equation}}

\begin{document}
\title{Hybrid phase-space--Fock-space approach to evolution of a driven nonlinear resonator}
\author{Mostafa Khezri}
\email[email: ]{mostafa.khezri@email.ucr.edu}
\affiliation{Department of Electrical and Computer Engineering, University of California, Riverside, California 92521, USA}
\affiliation{Department of Physics, University of California, Riverside, California 92521, USA}
\author{Alexander N. Korotkov}
\affiliation{Department of Electrical and Computer Engineering, University of California, Riverside, California 92521, USA}

\date{\today}

\begin{abstract}
We analyze the quantum evolution of a weakly nonlinear resonator due to a classical near-resonant drive and damping.
The resonator nonlinearity leads to squeezing and heating of the resonator state.
Using a hybrid phase-space--Fock-space representation for the resonator state within the Gaussian approximation, we derive evolution equations for the four parameters characterizing the Gaussian state.
Numerical solution of these four ordinary differential equations is much simpler and faster than simulation of the full density matrix evolution, while providing good accuracy for the system analysis during transients and in the steady state.
We show that steady-state squeezing of the resonator state is limited by 3\,dB; however, this limit can be exceeded during transients.
\end{abstract}
\maketitle

\section{Introduction}\label{sec:intro}

Nonlinear quantum oscillators have been a subject of various studies for a long time \cite{Dykman1973, Drummond1980, Tanas1984, Lugiato1984, Dykman-2012book}. The renewed interest in this system is caused by the wide use of microwave resonators in superconducting quantum computing circuits \cite{Wallraff2004, Siddiqi2005}, as well as reaching a quantum regime for nanomechanical resonators \cite{Blencowe2004, Schwab2005, O'Connell2010, Wollman2015}.
In particular, during dispersive measurement of superconducting qubits \cite{Blais2004, Wallraff2004, Chow2014, Jeffrey2014, Saira2014}, nonlinearity of the measurement resonator is induced by its coupling with the qubit; this nonlinearity causes significant deviations from the standard dispersive regime in the case of a moderately or strongly driven resonator \cite{Boissonneault2010, Reed2010, Khezri2016}. The nonlinearity of Josephson-junction-based resonators is used in experiments for near-quantum-limited microwave signal amplification \cite{Vijay2008, Bergeal2010, Mallet2011, Vijay2011}.

Driven nonlinear resonators can produce squeezed states \cite{Tanas1984, Lugiato1984, Tanas1989, Milburn1986, Kitagawa1986,Dykman-2012} (note that quantum squeezing is closely related to classical fluctuations, e.g., \cite{Tomita1974, Dykman1979, Ludwig1975}). Even though squeezed states are usually discussed for parametrically driven linear resonators \cite{Walls2008, Drummond2004} (in optics a nonlinear material can be used to produce a parametric drive at a doubled frequency), there is a similarity between these two systems \cite{Dykman2007, Vijay2008, laflamme2011}. In particular, it can be shown that a nonlinear resonator near the bifurcation point at large photon numbers is equivalent to a degenerate parametric amplifier driven with a detuned pump \cite{laflamme2011}. Squeezed states can be used to improve measurement accuracy \cite{Caves1981, Giovannetti2004} in a range of applications, such as gravitational wave detectors \cite{LIGO2013},  superconducting qubit readout \cite{Sete2013,Barzanjeh2014,Didier2015,Didier2015b,Govia2017,Eddins2017}, and nano/micromechanical position measurement \cite{Ruskov2005, Peano2015, Wollman2015}. There is currently a significant experimental interest in producing squeezed microwave states with Josephson parametric amplifiers \cite{Movshovich1990, Mallet2011, Eichler2011, Flurin2012, Murch2013, Toyli2016}; the self-developing squeezing due to  the nonlinearity of a microwave resonator (with revival and formation of ``cat'' states) has also been demonstrated experimentally \cite{Kirchmair2013}.

It is well known that the steady state of a parametrically driven resonator cannot be squeezed beyond 3 dB \cite{Milburn1981,Walls2008,Collett1984}; in other words, any (instantaneous) quadrature variance is not less than $1/2$ of the ground-state value (the 3 dB squeezing is reached in the ideal case at the threshold of parametric instability; note that for the narrow-band definition, squeezing in the same case is 6 dB \cite{Rugar1991}). This limit applies only to the resonator state (intracavity field), while squeezing  of the reflected field outside of the cavity is unlimited  \cite{Collett1984,Yurke1984}. Various theoretical ideas \cite{Rabl2004, Kronwald2013, Szorkovszky2011, Ruskov2005, Jahne2009, Vanner2011} (based on reservoir engineering, weak measurements, injection of squeezed light, etc.) have been proposed to overcome the 3 dB limit for a nanomechanical resonator; recently this limit has been exceeded experimentally \cite{Lei2016}.

Because of the similarity between nonlinear and parametrically driven resonators in their use as amplifiers \cite{Vijay2008,laflamme2011}, it can be expected that squeezing of   driven nonlinear resonators is also limited by 3 dB. However, we are not aware of papers, which discuss this limit explicitly (related works are, e.g., Refs.\ \cite{Drummond1980, Dykman1994, Buks2006, Almog2007, Serban2010, Dykman-2012}; note explicit results for steady-state quantum fluctuations in Refs.\ \cite{Drummond1980} and \cite{Dykman-2012}). As a side result of this paper, we will show that the steady-state squeezing of a coherently driven nonlinear resonator is indeed limited by 3 dB. We will also show that during transients the squeezing can exceed this limit.

Previous studies of quantum dynamics of coherently driven nonlinear oscillators have used a variety of theoretical methods, including stochastic differential equations, Fokker-Planck equation, generalized $P$-representation, linearization of evolution equation, formalism of quasienergies, etc.  Usually the transients are neglected and only the steady state is analyzed. Moreover, most of the research has been focused on the regimes close to bifurcation or within the bistability range, in particular, with the goals to analyze switching between the quasistable states and to analyze amplification properties near the bifurcation point. In this paper we are mainly interested in the opposite regime: far from the bifurcation and/or bistability, so that the effects of nonlinearity are not yet very strong. This regime is relevant to the measurement of superconducting qubits, in which the weak nonlinearity of the microwave resonator is induced by its interaction with the qubit. Nevertheless, this weak nonlinearity may lead to a significant self-developing squeezing of the microwave field \cite{Sete2013}, which affects qubit measurement fidelity. Another difference of our analysis from most of the previous studies is that we are mainly interested in transients, not the steady state. This is also motivated by the importance of transients in fast measurement of superconducting qubits. Even though our motivation mainly comes from the use of weakly nonlinear  microwave resonators for qubit measurement, our results are equally applicable to the quantum dynamics of driven nanomechanical resonators, which always show some nonlinearity \cite{Cleland2003}.

In this paper, we analyze the evolution of a coherently driven weakly nonlinear resonator using a hybrid phase-space--Fock-space approach \cite{Khezri2016}.
This approach is based on the observation that quantum state evolution due to nonlinearity can be easily described in Fock space, while the effect of the drive and dissipation for a linear resonator is well described in phase space. We show that for large photon numbers, a Gaussian state \cite{Gardiner2004} in phase space has also an approximately Gaussian form in Fock space, thus obtaining a rather simple conversion between the Fock-space and phase-space representations within the Gaussian-state approximation. The conversion equations are then used to derive reasonably simple first-order ordinary differential equations, describing state evolution due to drive, dissipation, and weak nonlinearity.

These evolution equations are for one complex and three real  parameters, which characterize the Gaussian state of the resonator. The complex parameter describes the center of the Gaussian state in the phase plane; its evolution is given by an essentially classical equation, which takes into account nonlinearity. The three real parameters are Fock-space parameters, which after conversion into the phase space correspond to the minimum and maximum quadrature variances (therefore to squeezing and ``unsqueezing'') and to the phase of the minimum-variance quadrature. The product of the minimum and maximum variances (ratio of unsqueezing and squeezing) corresponds to an effective temperature, which can be significantly higher \cite{Dykman-2012} than the bath temperature. We note that our approach is physically similar to linearization of fluctuations around the classical trajectory within the Gaussian approximation \cite{Ludwig1975}, even though it is based on a different framework.

After deriving the hybrid phase-Fock-space evolution equations, we numerically compare their results with the master (Lindblad) equation simulations. We find quite good accuracy, with an inaccuracy scaling inversely proportional to the number of photons in the system. Even though our approximation formally requires large number of photons, it still works well when the resonator evolution starts from the ground state. In our simulations with a few hundred photons in the system, the typical infidelity compared with the master equation simulations is about $10^{-3}-10^{-4}$, while being faster by a factor of over $10^{5}$ (fractions of a second instead of hours). Compared with the coherent-state approximation, our method for the simulated cases is more accurate by about a factor of $10^2$, which indicates the importance of taking into account self-developing squeezing and heating.

Thus, our main result in this paper is the derivation of relatively simple and computationally efficient equations, which describe the quantum evolution of a driven and damped weakly nonlinear resonator in the case of large photon numbers. As an example of using these equations, we derive the 3 dB squeezing limit discussed above for the steady state and numerically show that this limit can be exceeded during transients. Note that we analyze only the state of the resonator (intracavity field), while the analysis of the reflected field is left for future studies (the problem is that for a non-pulsed  propagating field, the standard definition of squeezing is applicable only in the steady state, so for transients we will need to modify the definition; analysis will probably require the use of either the input-output theory \cite{Gardiner1985, YurkeDenker1984, Gardiner2004, Clerk2010} for the linearized system or the approach of weak measurements \cite{Wiseman2010,Korotkov2016}).

The range of validity for our approach seems to be essentially the same as for validity of the Gaussian approximation. Note that for small number of photons in the resonator, the resonator is practically linear, while for large number of photons, the resonator is practically semi-classical, and in both cases the Gaussian approximation is applicable. This is why our approach works well in a rather wide range, except the vicinity of the bifurcation point, where unsqueezing becomes too large; also, within the bistability region our approach cannot describe gradual mixing of quasistable states, which corresponds to classical switching between them. We analyze the accuracy of our approach numerically, by comparing its results with results of simulations based on the master equation.

The paper is organized as follows.
In Sec.\ \ref{sec:problem} we describe the system and pose the problem.
In Sec.\ \ref{sec:Gaussian} we review the Gaussian states and corresponding phase-space evolution equations for a driven and damped linear resonator.
Then in Sec.\ \ref{subsec:FG} we introduce Fock-space Gaussian states and discuss their equivalence to the usual (phase-space) Gaussian states in the case of large photon numbers, with explicit conversion relations between parameters of the phase-space and Fock-space representations. Using these conversion relations, in Sec.\ \ref{subsec:evol} we combine the Fock-space evolution due to nonlinearity with the phase-space evolution due to drive and damping, thus deriving the hybrid phase-Fock-space evolution equations, which are the main result of this paper.
Section \ref{sec:numerical} is devoted to analysis of the numerical accuracy of our approach. We start with calculating the fidelity of the conversion between the Gaussian and Fock-space Gaussian states in Sec.\ \ref{subsec:conversion-fidelity}, and then in Sec.\ \ref{subsec:fid-hybrid} we compare results of the hybrid evolution equations with the master equation simulations.
In Sec.\ \ref{sec:3dB} the hybrid evolution equations are used to show that steady-state squeezing of the resonator state is limited by 3 dB,
and it is also shown numerically that squeezing during transients can exceed the 3 dB limit. We conclude in Sec.\ \ref{sec:conclusion}.
In Appendix \ref{app:rf} we discuss derivation of the Gaussian state evolution equations for a linear resonator under coherent  drive and damping. In Appendix \ref{app:conversion} we show that at large photon numbers, a Fock-space Gaussian state can be approximated by a phase-space Gaussian state, and derive the corresponding conversion relations. Appendix \ref{app:steady} discusses analytical results for squeezing in the steady state.

\section{System and problem}\label{sec:problem}

We analyze the quantum state evolution of a weakly nonlinear resonator, which is coherently (classically) driven at frequency $\omega_{\rm d}$ and damped due to energy relaxation with rate $\kappa$  at bath temperature $T_{\rm b}$.
The goal is to find a reasonably simple approximate description of this evolution, suitable for large number of photons in the resonator (we will use the terminology of photons, though for a mechanical resonator the terminology of phonons would be more appropriate).

Without damping, the laboratory-frame Hamiltonian of the considered system is ($\hbar=1$)
    \begin{eqnarray}
&& H_{\rm lf} = H^{\rm lf}_{\rm r} +H^{\rm lf}_{\rm d},
    \\
&& H_{\rm r}^{\rm lf} = \sum_n E(n) \, \ket{n}\bra{n}, \,\,\,\, E(n)=\sum_{k=0}^{n-1}\omega_{\rm r}(k),
    \label{H-lf-r}\\
&&
    H^{\rm lf}_{\rm d} = 2{\rm Re} [\varepsilon (t)\, e^{-i\omega_{\rm d}t}] \, (a^\dagger +a),
    \label{H-lf-d}
    \end{eqnarray}
where $|n\rangle$ is $n$th eigenstate of the resonator, with corresponding eigenenergy $E(n)$ expressed via the resonator frequency $\omega_{\rm r}(n)=E(n+1)-E(n)$, which slightly changes with the level number [we use $E(0)=0$], $\varepsilon(t)$ is the complex amplitude of the drive at frequency $\omega_{\rm d}$, and $a=\hat{x}+i\hat{p}$ is the annihilation operator, while $a^\dagger=\hat{x}-i\hat{p}$ is the creation operator.
Here $\hat{x}$ and $\hat{p}$ are normalized position and momentum operators, $\hat{x}=\hat{X}\sqrt{m\omega_{\rm r0}/2}$ and $\hat{p}=\hat{P}/\sqrt{2m\omega_{\rm r0}}$, where $\hat{X}$ and $\hat{P}$ are actual position and momentum operators, $m$ is effective mass, and in the normalization we use $\omega_{\rm r0}\equiv\omega_{\rm r}(0)$; however, this particular value is not important, since we assume a weak nonlinearity, $|\omega_{\rm r}(n)-\omega_{\rm r}(0)|\ll \omega_{\rm r} (0)$.
The assumption of weak nonlinearity also allows us to use the standard matrix elements for the annihilation operators, $\langle k|a|n\rangle=\sqrt{n}\, \delta_{n-1,k}$.
Note that for a linear resonator, $\omega_{\rm r}(n)=\omega_{\rm r0}$, the Hamiltonian (\ref{H-lf-r}) reduces to the standard form $H_{\rm r}^{\rm lf} =\omega_{\rm r0}a^\dagger a$.
Within the rotating wave approximation (RWA), the drive Hamiltonian (\ref{H-lf-d}) becomes
$H^{\rm lf}_{\rm d} = \varepsilon (t)\, e^{-i \omega_{\rm d} t} a^\dagger + \varepsilon^*(t)\, e^{i \omega_{\rm d} t} a$.
The RWA is natural for a weakly nonlinear resonator and near-resonant drive, $|\omega_{\rm d}-\omega_{\rm r}(n)|\ll \omega_{\rm d}$.
In some cases RWA misses experimentally important effects  \cite{Sank2016}; however, it should be sufficient for the simple system we consider here.

In the rotating frame based on the drive frequency $\omega_{\rm d}$, the RWA Hamiltonian becomes
$H_{\rm rf} = H^{\rm rf}_{\rm r} +H^{\rm rf}_{\rm d}$ with
    \begin{eqnarray}
&& H_{\rm r}^{\rm rf} = \sum_n E_{\rm rf}(n) \, \ket{n}\bra{n}, \,\,\, E_{\rm rf}(n)=\sum_{k=0}^{n-1}[\omega_{\rm r}(k)-\omega_{\rm d}],\quad
    \label{eq:H-r}\\
&&
    H^{\rm rf}_{\rm d}=  \varepsilon (t)\, a^\dagger + \varepsilon^*(t)\,  a .
    \label{H-rf-d} \end{eqnarray}
In this paper we will mostly use the rotating frame.

The evolution of the system density matrix $\rho$ due to Hamiltonian $H$ (in either laboratory or rotating frame) and energy relaxation with rate $\kappa$ is described by the standard master equation in the Lindblad  form \cite{Zeldovich1969, Walls2008, Weiss2012},
    \begin{align}\label{eq:master}
	\dot{\rho} =&\, i[\rho,H] + \kappa(n_{\rm b}+1)(a\rho a^\dagger-a^\dagger a\rho/2-\rho a^\dagger a/2) \nonumber \\
	&+ \kappa\, n_{\rm b}(a^\dagger\rho a-a a^\dagger \rho/2-\rho a a^\dagger/2),
    \end{align}
where
    \begin{equation} n_{\rm b}=\frac{1}{e^{\omega_{\rm r0}/T_{\rm b}}-1}=\frac{\coth(\omega_{\rm r0}/2T_{\rm b})-1}{2}
    \label{n-b-def}\end{equation} is the average number of thermal photons for the bath temperature $T_{\rm b}$.
Note that the evolution equation (\ref{eq:master}) is generally not correct for a nonlinear resonator (e.g., Appendix B4 of \cite{Korotkov2013}); however, we use it, assuming a weak nonlinearity.
The problem with applicability of the Lindblad equation (\ref{eq:master}) stems from the fact that it requires indistinguishability of the emitted and/or absorbed photons \cite{Korotkov2013}. However, for a weakly nonlinear resonator, the photons emitted from (absorbed by) different levels have slightly different frequencies and can be distinguished spectroscopically if the frequency difference exceeds the level width.
To estimate the effect, let us assume that unsqueezing is not too large, so the typical number of photons is $\bar{n}\pm\sqrt{\bar{n}}$, where $\bar{n}$ is the average photon number. Then the frequency difference is about $\sqrt{\bar{n}} \, (d\omega_{\rm r}/dn)$, while the level width is approximately $\kappa \bar{n}$. Therefore, indistinguishability requires $\bar{n}\gg \kappa^{-2}(d\omega_{\rm r}/dn)^2$. For our typical parameters used in Sec.\ \ref{sec:numerical}, the nonlinearity is quite small, so that  $\kappa^{-2}(d\omega_{\rm r}/dn)^2 \sim 10^{-5}$;  therefore, the indistinguishability condition is well satisfied and
the Lindblad equation (\ref{eq:master}) is accurate.

Solving Eq.\ (\ref{eq:master}) numerically in the Fock space, we can find the resonator state evolution.
However, for over $\sim$100 average photons in the resonator the numerical solution becomes slow, and for over $\sim$500 photons it becomes computationally intractable on a personal computer because of too large Hilbert space.
Note that over 500 photons in the  resonator can be used for a dispersive measurement of a superconducting qubit \cite{Sank2016,Bultink2016}.

In this paper, we develop an approach which permits a simple analysis of evolution at this large number of photons.
To a significant extent, the approach is based on the observation that evolution of a {\it linear} resonator can be described by Gaussian states in many situations \cite{Halliwell1995}.
Using the fact that a weak nonlinearity keeps the evolving state Gaussian (in the leading order), we will find the corresponding evolution equations.
This greatly simplifies analysis, since a Gaussian state is characterized by only 5 real parameters, instead of $N^2$ parameters for a density matrix involving up to $N$ Fock states.

We will first review Gaussian states and evolution of a driven linear resonator, and then will show how a Gaussian state can be approximately converted into a Fock-space state, for which it is easy to introduce evolution due to nonlinearity.

\section{Evolution of a linear resonator} \label{sec:Gaussian}

Without nonlinearity, a Gaussian initial state remains Gaussian during evolution, while initially non-Gaussian state gradually becomes Gaussian \cite{Zurek1993,Halliwell1995}.
In this section we briefly review properties of the Gaussian states and discuss evolution of a linear resonator state due to applied drive and damping.

\subsection{Brief review of Gaussian states}\label{subsec:Gaussian-review}

Gaussian states \cite{Gardiner2004, Weedbrook2012, Braunstein2005, Ferraro2005,*Ferraro2005arxiv} are defined as states for which the Wigner function \cite{Gardiner2004,Gerry2005} has a Gaussian form (generally with an arbitrary number of dimensions).
For a one-dimensional (single-mode) system with position operator $\hat{X}$ and conjugate momentum operator $\hat{P}$, the Wigner function of a Gaussian state is
    \begin{align}\label{eq:W}
	\mathcal{W}(X,P) = \frac{\exp\left(-\frac{1}{2}\vec{V}^T\pmb{D}^{-1}\vec{V}\right)} {2\pi\sqrt{\text{Det}(\pmb{D})}}
    \end{align}
where $\vec{V}=(X-X_{\rm c} , P-P_{\rm c})^T$, $X_{\rm c}=\mean{\hat{X}}$, $P_{\rm c}=\mean{\hat{P}}$, and elements of the covariance matrix $\pmb{D}$ are $D_{11}= D_X= \mean{\hat{X}^2}-\mean{\hat{X}}^2$, $D_{22}=D_P=\mean{\hat{P}^2}-\mean{\hat{P}}^2$, and $D_{12}=D_{21}=D_{XP}= \mean{\hat{X}\hat{P}+\hat{P}\hat{X}}/2 -\mean{\hat{X}} \mean{\hat{P}}$.
The Husimi $Q$-function, Glauber-Sudarshan $P$-function and density matrix (in $X$ or $P$ space) of a Gaussian state have a Gaussian form as well \cite{Gardiner2004,Marian1993}.

For a linear resonator with Hamiltonian $H_{\rm r}^{\rm lf}=\omega_{\rm r}a^\dagger a$ (constant frequency $\omega_{\rm r}$), we can introduce the dimensionless (normalized) operators of position and momentum in the standard way as $\hat{x} = \hat{X}/(2\sigma_{x,\rm gr})$ and $\hat{p} = \hat{P}/(2\sigma_{p,\rm gr})$, where $\sigma_{x,\rm gr}$ and  $\sigma_{p,\rm gr}$ are the standard deviations of the ground state in the position and momentum representations, so that $\hat{x}=(a+a^\dagger)/2$ and $\hat{p}=(a-a^\dagger)/2i$.
For the normalized operators, the Wigner function $W(x,p)$ has exactly the same form as Eq. (\ref{eq:W}), except now  $\vec{V}=(x-x_{\rm c} , p-p_{\rm c})^T$, $x_{\rm c}=\mean{\hat{x}}$, $p_{\rm c}=\mean{\hat{p}}$, and elements of the covariance matrix are now $D_{11}= D_x= \mean{\hat{x}^2}-\mean{\hat{x}}^2$, $D_{22}=D_p=\mean{\hat{p}^2}-\mean{\hat{p}}^2$, and $D_{12}=D_{21}=D_{xp}= \mean{\hat{x}\hat{p}+\hat{p}\hat{x}}/2 -\mean{\hat{x}}\mean{\hat{p}}$.
Explicit form of the Wigner function for a Gaussian state is
    \begin{align}\label{eq:W-norm}
& \hspace{-0.1cm}  W(x,p) = \left( 2\pi\sqrt{D_xD_p-D_{xp}^2} \, \right)^{-1}
    \nonumber \\
	& \hspace{0.3cm} \times \exp\bigg[- \frac{D_p(\Delta x)^2 + D_x(\Delta p)^2 - 2D_{xp} \Delta x \Delta p )}{2(D_xD_p-D_{xp}^2)}\bigg],
    \end{align}
where $\Delta x=x-x_{\rm c}$ and $\Delta p=p-p_{\rm c}$.
The Wigner functions (\ref{eq:W}) and (\ref{eq:W-norm}) are normalized as $\int \mathcal{W}(X,P)\, dX\, dP = \int W(x,p)\, dx\,dp = 1$.

With the quadrature operator along direction $\varphi$ defined as
\begin{equation}\label{x-varphi}
	\hat{x}_\varphi \equiv \frac{a e^{-i\varphi} + a^\dagger e^{i\varphi}}{2} =\hat{x} \cos\varphi +\hat{p} \sin\varphi,
\end{equation}
the variance $\sigma^2_{x_\varphi} \equiv \langle \hat{x}_\varphi^2\rangle -\langle \hat{x}_\varphi\rangle^2$ of this quadrature for the Gaussian state is
\begin{equation}\label{eq:quad-var}
	\sigma^2_{x_\varphi} = D_x\cos^2\varphi + D_p\sin^2\varphi + 2D_{xp}\cos\varphi\sin\varphi.
\end{equation}

\begin{figure}[t]
	\begin{center}
\includegraphics[width=5cm]{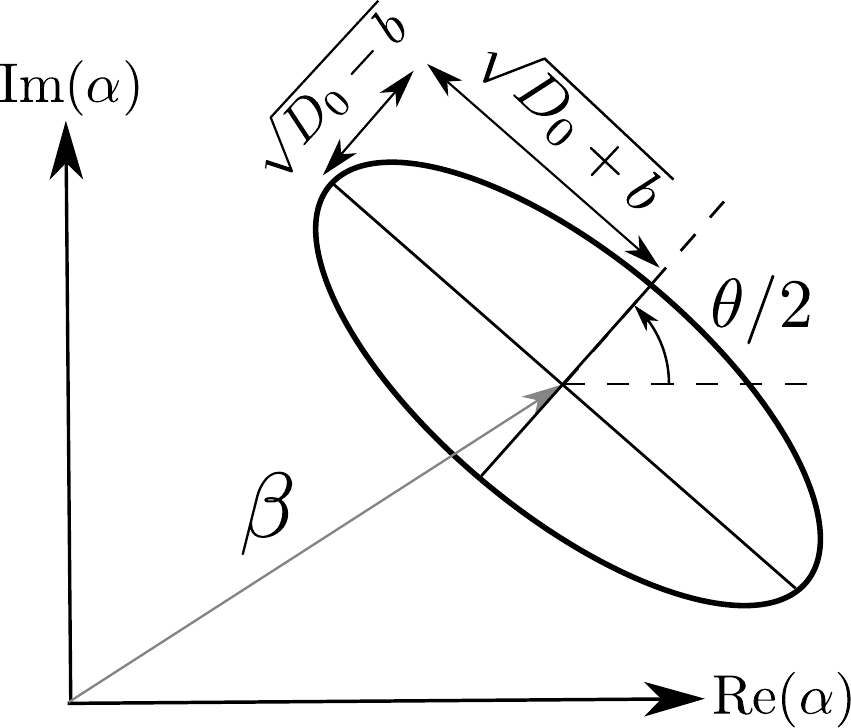}
    \end{center}
    \caption{
    Phase-space illustration of a Gaussian state.
    The ellipse corresponds to one standard deviation for the quadrature operators along any direction.
    It is also the  contour line for the Wigner function being a factor $\sqrt{e}$ less than its maximum value.
    The ellipse center has coordinates $(x_{\rm c}, p_{\rm c})$, which on the complex plane correspond to $\langle a\rangle =x_{\rm c}+ip_{\rm c}$. The minimum and maximum quadrature variances are $D_0-b$ and $D_0+b$, respectively. The minimum-variance-direction angle is $\Theta/2$. In the rotating frame we use notation $\theta$ instead of $\Theta$.
    }\label{fig:quad}
\end{figure}

Let us introduce real variables $D_0>0$ and $b\geq 0$ as
\begin{equation}\label{eq:D-b}
	D_0\equiv \frac{D_x+D_p}{2}, \,\,\, b^2\equiv \frac{(D_x-D_p)^2}{4} + D_{xp}^2,
\end{equation}
then the quadrature variance (\ref{eq:quad-var}) can be rewritten as
\begin{eqnarray}\label{eq:quad-var-D0-b}
&& \sigma^2_{x_\varphi} = D_0 -b\cos (2\varphi-\Theta),
    \\
&& \Theta = {\rm arctan}\left( \frac{2D_{xp}}{D_x-D_p}\right)
+\frac{\pi}{2} [1+{\rm sign} (D_x-D_p)] .   \quad
    \label{eq:Theta-1}
\end{eqnarray}
Equation \eqref{eq:quad-var-D0-b} shows that $D_0-b$ and $D_0+b$ are the minimum and maximum quadrature variances respectively, and the direction of the minimum quadrature makes the angle $\Theta/2$ with the $x$-axis (see Fig.\,\ref{fig:quad}). Note that
    \begin{equation}
(D_0+b)(D_0-b)=D_xD_p-D_{xp}^2.
    \end{equation}

The Wigner function in the rotated ``diagonal basis'' with $x^{\rm d}$ being the coordinate along the {\it minimum} quadrature is
    \begin{align}\label{eq:wigner-eigen}
	W(x^{\rm d},p^{\rm d})=&\left(2\pi\sqrt{(D_0-b)(D_0+b)}\right)^{-1} \nonumber \\
& \times \exp\left[-\frac{(x^{\rm d}-x^{\rm d}_{\rm c})^2}{2(D_0-b)}-\frac{(p^{\rm d}-p^{\rm d}_{\rm c})^2}{2(D_0+b)}\right],
    \end{align}
where $x^{\rm d}+ip^{\rm d}=(x+ip)\, e^{-i\Theta/2}$ and similarly  $x^{\rm d}_{\rm c}+ip^{\rm d}_{\rm c}=(x_{\rm c}+ip_{\rm c})\, e^{-i\Theta/2}$.
This formula shows that the contour lines for the Wigner function in the phase space of $x$ and $p$ are ellipses (Fig.\ \ref{fig:quad}).

The Husimi $Q$-function \cite{Walls2008} for the Gaussian state can be obtained using the standard relation $Q(x,p)=\frac{2}{\pi}\int W(x',p')\, e^{-2[ (x-x')^2 +(p-p')^2]} \, dx'\, dp'$.
In particular, in the diagonal basis we find
    \begin{align}\label{eq:Q-gaussian-eigen}
& \hspace{-0.2cm} Q(x^{\rm d},p^{\rm d})=\left(2\pi\sqrt{(D_0-b+1/4)(D_0+b+1/4)}\right)^{-1}  \qquad
    \nonumber \\
& \hspace{0.3cm} \times \exp\left[-\frac{(x^{\rm d}-x^{\rm d}_{\rm c})^2}{2(D_0-b+1/4)}-\frac{(p^{\rm d}-p^{\rm d}_{\rm c})^2}{2(D_0+b+1/4)}\right].
    \end{align}
We see that the $Q$-function (\ref{eq:Q-gaussian-eigen}) has the same Gaussian form as the Wigner function (\ref{eq:wigner-eigen}), but variances for the both axes are increased by $1/4$.

It is useful to write the Gaussian state parameters in terms of average values of the operators $a$, $a^2$, and $a^\dagger a$,
    \begin{align}
	& D_0 = \frac{1}{2}\Big[ \mean{a^\dagger a} + \frac{1}{2} - (\text{Re}\mean{a})^2 - (\text{Im}\mean{a})^2\Big] , \label{eq:a-to-D}\\
	& b = \frac{1}{2}\Big[ \left[\text{Re}\mean{a^2}-(\text{Re}\mean{a})^2+ (\text{Im}\mean{a})^2\right]^2
    \nonumber \\
	& \hspace{1.0cm} + \left( \text{Im}\mean{a^2}-2\text{Re}\mean{a}\,\text{Im}\mean{a}\right)^2 \Big]^{1/2} \, \,
    \\
	& \Theta = \arctan\left(\frac{\text{Im}\mean{a^2}-2\text{Re}\mean{a}\,
\text{Im}\mean{a}}
{\text{Re}\mean{a^2}-(\text{Re}\mean{a})^2+(\text{Im}\mean{a})^2}
\right)
    \nonumber \\
& \hspace{0.5cm} +\frac{\pi}{2} \{ 1+{\rm sign} [ \text{Re}\mean{a^2}
-(\text{Re}\mean{a})^2+(\text{Im}\mean{a})^2 ]\} ,
    \label{eq:a-to-theta} \\
& x_{\rm c}+ip_{\rm c} = \mean{a}.
    \end{align}

Besides introducing the Gaussian states via the Wigner function, it is also possible to introduce them as displaced squeezed thermal states (DSTS) \cite{Fearn1988,Kim1989,Marian1993}, so that the density matrix is
    \begin{equation}\label{eq:DSTS}
	\rho_\text{DSTS} = D(\alpha)\, S(\xi)\, \nu_{n_\text{th}}\, S(\xi)^\dagger \, D(\alpha)^\dagger,
    \end{equation}
where $\alpha=\langle a\rangle=x_{\rm c}+ip_{\rm c}$ is the phase-plane state center, $D(\alpha) = \exp(\alpha a^\dagger - \alpha^* a)$ is the displacement operator, $S(\xi)=\exp[\frac{1}{2}\xi^*a^2 - \frac{1}{2}\xi (a^\dagger)^2]$ is the squeezing operator with squeezing parameter $\xi=re^{i\Theta}$ (the angle $\Theta/2$ determines the short axis direction and therefore $\Theta$ is the same as discussed above), and $\nu_{n_\text{th}}$ is the thermal state, defined as
    \begin{equation}
	\nu_{n_{\text{th}}} = \frac{1}{1+n_{\text{th}}}\sum_{k=0}^\infty \left( \frac{n_{\text{th}}}{1+n_{\text{th}}} \right)^k \ket{k}\bra{k},
    \label{nu-thermal}
    \end{equation}
where $|k\rangle$ is $k$th Fock state and $n_\text{th}=\text{Tr}(a^\dagger a \,\nu_{n_{\text{th}}})$ is the average number of thermal photons.
Note that Eq.\ (\ref{nu-thermal}) describes an equilibrium state of a linear resonator at finite temperature without drive, and in that case $n_{\rm th}$ is equal to the thermal photon number for the bath, $n_{\rm b}$, given by Eq.\ (\ref{n-b-def}).
However, in the non-equilibrium case considered in this paper,  $n_{\rm th}$ is not equal to $n_{\rm b}$.
It is still possible to define an {\it effective} temperature $T_{\rm eff}$ for a Gaussian state (\ref{eq:DSTS}) via the same relation,
    \begin{equation}
\coth(\omega_{\rm r}/2T_{\rm eff})= 1+ 2 n_{\rm th}.
    \end{equation}
Note that the average photon number $\bar{n}$ for a Gaussian state has a contribution proportional (but not equal) to $n_{\rm th}$,
    \begin{equation}\label{eq:nbar}
	\bar{n}=\text{Tr}(a^\dagger a\,\rho_\text{DSTS}) = |\alpha|^2 + (1+2n_{\rm th}) \sinh^2 r + n_\text{th},
    \end{equation}
while from Eq.\ (\ref{eq:a-to-D}) we find a simple expression
    \begin{equation}
\bar{n} =  |\alpha|^2 + 2D_0 -1/2 .
    \label{nbar-via-D} \end{equation}

To relate parameters $r$ and $n_{\rm th}$ of the DSTS state to the parameters of the Gaussian state (\ref{eq:W-norm}), we can calculate averages $\langle a\rangle$, $\langle a^2\rangle$, and $\langle a^\dagger a\rangle$ for the state (\ref{eq:DSTS}), and use these results to find the variances
    \begin{align}
	&D_x = (1/4+ n_{\text{th}}/2)(\cosh 2r - \sinh 2r\cos \Theta ), \label{eq:D-to-r-i} \\
	&D_p = (1/4+ n_{\text{th}}/2)(\cosh 2r + \sinh 2r\cos \Theta ), \\
	&D_{xp} = -(1/4+ n_{\text{th}}/2)\sinh 2r\sin \Theta . \label{eq:D-to-r-f}
    \end{align}
Comparing Eqs.\,\eqref{eq:D-to-r-i}--\eqref{eq:D-to-r-f} with Eqs.\,\eqref{eq:D-b}--(\ref{eq:Theta-1}), we find the equivalence for
    \begin{equation}\label{eq:nth-r}
	n_\text{th} = 2\sqrt{(D_0+b)(D_0-b)} - \frac{1}{2}, \,\,\,\,
	\tanh 2r = \frac{b}{D_0},
    \end{equation}
and the same angle $\Theta$.

As follows from the discussion above, a Gaussian state is determined by five real parameters.
Two parameters,  $x_{\rm c}$ and $p_{\rm c}$, define the state center on the phase plane; it is convenient to use their complex combination $\alpha=x_{\rm c}+ip_{\rm c}$.
Three real parameters define the ``shape'' (see Fig.\ \ref{fig:quad}), which can be characterized either by $D_x$, $D_p$, and $D_{xp}$ or by $D_0$, $b$, and $\Theta$ or by $r$, $\Theta$, and $n_{\rm th}$.
A Gaussian state is in general a mixed state.
A pure Gaussian state is a minimum-uncertainty squeezed state, characterized by 4 real parameters; for such a state $D_xD_p-D_{xp}^2=(D_0-b)(D_0+b)=1/16$ and $n_{\rm th}=0$.
A coherent state is characterized by only 2 real parameters, which define the center; then $D_x=D_p=D_0=1/4$, $D_{xp}=b=n_{\rm th}=0$, and $\Theta$ is not important.

Note that our discussion in this section used the laboratory frame.
In this frame, the evolution due to Hamiltonian $H_{\rm r}^{\rm lf}=\omega_{\rm r}a^\dagger a$ (in the absence of drive and damping) rotates the state center in Fig.\ \ref{fig:quad} {\it clockwise} with angular velocity $\omega_{\rm r}$.
Moreover, the whole phase-space picture in Fig.\ \ref{fig:quad} rotates clockwise with $\omega_{\rm r}$.
This means that parameters $D_0$ and $b$ do not change with time, while
the angle $\Theta /2$ evolves as $d(\Theta/2)/dt=-\omega_{\rm r}$, and therefore $\dot{\Theta}=-2 \omega_{\rm r}$.
Since $D_0$ and $b$ do not change, the parameters $r$ and $n_{\rm th}$ are also constant -- see Eq.\ (\ref{eq:nth-r}).
In the rotating frame based on the frequency $\omega_{\rm d}$, the picture in Fig.\ \ref{fig:quad} additionally rotates  counterclockwise with angular velocity $\omega_{\rm d}$, so that the net evolution is clockwise rotation with angular velocity $\omega_{\rm r}-\omega_{\rm d}$.
Thus, in the rotating frame, the parameters $D_0$, $b$, $r$, and $n_{\rm th}$ are the same as in the laboratory frame, while the rotating-frame angle parameter $\theta$ is related to $\Theta$ as
    \begin{equation}\label{Theta-theta}
	\theta =\Theta + 2\omega_{\rm d} t,
    \end{equation}
and it evolves as $\dot{\theta}=-2 (\omega_{\rm r}-\omega_{\rm d})$.
Descriptions of the Gaussian states in the rotating and laboratory frames are practically the same, except $\Theta$ is replaced with $\theta$ and $\omega_{\rm r}$ is replaced with $\omega_{\rm r}-\omega_{\rm d}$, as expected for the rotating-frame Hamiltonian $H_{\rm r}^{\rm rf}=(\omega_{\rm r}-\omega_{\rm d})\,a^\dagger a$.
Note, however, that the conversion between the actual position and momentum operators ($\hat {X}$, $\hat{P}$) and the corresponding normalized operators ($\hat {x}$, $\hat{p}$) should still be based on the actual frequency $\omega_{\rm r}$ and not on $\omega_{\rm r}-\omega_{\rm d}$.
The relation between the laboratory frame and the rotating frame is discussed in more detail in the Appendix \ref{app:rf}.
Evolution in the presence of drive and damping is discussed next.

\subsection{Evolution equations}

For a linear harmonic oscillator with $H_{\rm r}^{\rm lf}=\omega_{\rm r}a^\dagger a$, the evolution  (\ref{eq:master}) due to drive (\ref{H-lf-d}) and damping $\kappa$ at bath temperature $T_{\rm b}$, preserves state Gaussianity and leads to the following evolution equations in the {\it laboratory frame} \cite{Halliwell1995, Doherty1999, Hopkins2003, Ruskov2005},
    \begin{align}
&\dot{x}_{\rm c} = \omega_{\rm r} p_{\rm c},
    \label{eq:evol-x} \\
&\dot{p}_{\rm c} = -\omega_{\rm r} x_{\rm c} -\kappa p_{\rm c} - 2\text{Re}(\varepsilon e^{-i\omega_{\rm d}t}),
    \label{eq:evol-p} \\
&\dot{D}_x = 2\omega_{\rm r} D_{xp},
    \label{eq:evol-Dx} \\
&\dot{D}_p = -2\omega_{\rm r} D_{xp} -2\kappa D_p +(\kappa /2)\coth (\omega_{\rm r}/2T_{\rm b}) ,
    \label{eq:evol-Dp} \\
&\dot{D}_{xp} = -\omega_{\rm r}(D_x-D_p) -\kappa D_{xp}.  \label{eq:evol-Dxp}
    \end{align}
Note that the evolution of the state center ($x_{\rm c}$ and $p_{\rm c}$) is decoupled from the evolution of the variances, and the drive $\varepsilon$ contributes only to $\dot{p}_{\rm c}$ (as a classical force).
The state center oscillates with the resonator frequency $\omega_{\rm r}$ (intrinsically, neglecting effects of $\kappa$ and $\varepsilon$), while the variances oscillate with doubled frequency, $2\omega_{\rm r}$.
Also note that Eqs.\ (\ref{eq:evol-x})--(\ref{eq:evol-Dxp}) do not rely on the RWA assumption.

Using the RWA (which symmetrizes coordinates $x$ and $p$) and going into the {\it rotating frame} based on the drive frequency $\omega_{\rm d}$, so that the Gaussian state center is characterized by a slowly changing complex number $\beta$ in the standard phase space,
    \begin{equation}
	\beta = (x_{\rm c} + ip_{\rm c})\, e^{i\omega_{\rm d}t},
    \end{equation}
from Eqs.\ (\ref{eq:evol-x})--(\ref{eq:evol-Dxp}) we can derive (see Appendix\,\ref{app:rf}) the following evolution equations \cite{Zeldovich1969, Paris2003} (see also \cite{Serafini2005,Marian1993}) for the parameters $\beta$, $D_0$, $b$, and $\theta$,
    \begin{align}
&\dot{\beta} = -i(\omega_{\rm r}-\omega_{\rm d})\beta -(\kappa/2)\beta -i\varepsilon,
    \label{eq:evol-beta} \\
&\dot{D_0} = -\kappa D_0 + (\kappa /4) \coth (\omega_{\rm r}/2T_{\rm b}) ,
    \label{eq:evol-D0} \\
&\dot{b} = -\kappa b,
    \label{eq:evol-b} \\
&\dot{\theta} = -2(\omega_{\rm r}-\omega_{\rm d}).
    \label{eq:evol-theta}
    \end{align}
Note that the drive does not affect evolution of the diagonal-basis variances $D_0\pm b$; however, the short-axis direction $\theta/2$ rotates clockwise with the detuning frequency $\omega_{\rm r}-\omega_{\rm d}$, similar to the rotation of the state center.

Equations \eqref{eq:evol-beta}--\eqref{eq:evol-theta} are the starting point of our analysis.
They describe evolution of a linear resonator using the phase-space language.
However, to include nonlinearity, we will need to approximately convert them into the Fock-space representation.
From now on, we will use only the rotating frame.

\section{Evolution of a weakly nonlinear resonator}

\subsection{Fock-space Gaussian state}\label{subsec:FG}

Generalizing the idea of Ref.\,\cite{Khezri2016}, let us introduce a state, for which the density matrix in the basis of eigenstates $|n\rangle$ (Fock space) has the following form,
    \begin{align}\label{eq:FG}
\rho_{mn} & = \frac{1}{\sqrt{2\pi W_1 |\beta|^2}}\exp\left[ -\frac{(\frac{n+m}{2}-|\beta|^2)^2}{2W_1|\beta|^2}-\frac{(n-m)^2} {8W_2|\beta|^2}\right]
    \nonumber \\
& \hspace{-0.3cm} \times \exp \bigg[ i\phi_\beta(n-m) -i\frac{2K}{|\beta|^2}\Big( \frac{n+m}{2}-|\beta|^2\Big) (n-m) \bigg] .
    \end{align}
We call it a Fock-space Gaussian state (because of quadratic dependence on $n$ and $m$ inside exponents) or, following the terminology of Ref.\ \cite{Khezri2016}, a sheared Gaussian state (because of a shearing effect produced by the $K$-term in the phase space).
The state (\ref{eq:FG}) is characterized by five real parameters: $|\beta|$, $\phi_\beta$, $W_1$, $W_2$, and $K$. Note that a physical $\rho_{mn}$ requires
    \be
    0<W_2\leq W_1.
    \ee

As shown in the Appendix \ref{app:conversion}, in the case $|\beta| \gg 1$ (while $W_1$, $W_2$, and $K$ are on the order of unity) this state is {\it approximately equal} to the standard Gaussian state discussed in Sec.\ \ref{sec:Gaussian}, so that
    \begin{equation}\label{eq:conv-beta}
	\beta = e^{i\phi_\beta} |\beta|
    \end{equation}
is (approximately) the state center, while the (approximate) conversion relations for the parameters $D_0$, $b$, and $\theta$ are
    \begin{align}
&D_0 = \frac{1}{8} \left[ \frac{1}{W_2} + W_1 (1+16K^2)\right],
    \label{eq:conv-D} \\
&b = \sqrt{ D_0^2 - W_1/(16 W_2) },
    \label{eq:conv-b} \\
&\theta = 2\phi_\beta + \arctan\Big( \frac{KW_1}{D_0-W_1/4} \Big)
    \nonumber \\
& \hspace{0.5cm} + (\pi /2)\,[1-{\rm sign}(D_0-W_1/4)] .
    \label{eq:conv-theta}
    \end{align}
The conversion becomes exact for $|\beta| \to \infty$.

While in the leading order $\langle a\rangle=e^{i\phi_\beta}|\beta|$ for the Fock-space Gaussian state (\ref{eq:FG}), more accurate calculations show the next-order correction proportional to $|\beta|^{-1}$,
    \begin{equation}\label{eq:a-aver-FG}
	\langle a\rangle = e^{i\phi_{\beta}} \left[ |\beta| - \frac{W_1+1/W_2-2}{8|\beta|}-\frac{2K^2 W_1}{|\beta|}-i \frac{KW_1}{|\beta|} \right] .
    \end{equation}
The overlap fidelity between the Gaussian and Fock-space Gaussian states becomes somewhat better if this correction is taken into account, so that a slightly shifted center corresponds to the same $\langle a\rangle$ for the Gaussian and Fock-space Gaussian states (see numerical results in Sec.\ \ref{subsec:conversion-fidelity}).
However, for simplicity we will not use the center correction (\ref{eq:a-aver-FG}) unless specifically mentioned.

Note that the trace of the state \eqref{eq:FG} is not exactly 1; however, the difference is negligible (exponentially small) for $|\beta| \gg 1$.
The Fock-space Gaussian state \eqref{eq:FG} is in general mixed;  it becomes pure if $W_2=W_1$, and in this case it reduces to the sheared Gaussian state introduced in Ref.\,\cite{Khezri2016}.
[Note a misprint in Eq.\,(33) of Ref.\,\cite{Khezri2016}, where the last exponent should actually be $-iK(n-|\beta|^2)^2/|\beta|^2$.]
Comparing Eqs.\ (\ref{eq:conv-D}) and (\ref{eq:conv-b}) with Eq.\ (\ref{eq:nth-r}), we find a useful relation for the thermal photon number,
    \begin{equation}
	n_{\rm th}=(\sqrt{W_1/W_2}-1)/2,
    \label{nth-W1/W2}\end{equation}
which is equivalent to the relation
    \begin{equation}
W_1/W_2=\coth ^2 (\omega_{\rm r}/2T_{\rm eff}) = 16(D_0+b)(D_0-b).
    \label{W1/W2} \end{equation}
Note that the ratio of the variances, $(D_0+b)/(D_0-b)$, and the angle $\theta /2-\phi_\beta$ are both functions of only two parameters: $K$ and $W_1 W_2$.

The quadrature variance $\sigma_{x_\varphi}^2$ along a direction $\varphi$ for the state \eqref{eq:FG} can be calculated as $\sigma_{x_\varphi}^2=D_0-b \cos (2\varphi -\theta)$ from Eqs.\ (\ref{eq:conv-D})--(\ref{eq:conv-theta}). In particular, for the direction along $\beta$ ($\varphi=\phi_\beta$) we find the variance $\sigma_{x_\varphi}^2=W_1/4$, while for the orthogonal direction ($\varphi=\phi_\beta+\pi/2$) we find the variance $\sigma_{x_\varphi}^2=1/(4W_2)+4K^2W_1$.

As follows from Eq.\ (\ref{eq:conv-theta}), in the case $K=0$,  the short axis (minimum variance) is either along the direction of $\beta$ ($\theta/2=\phi_\beta$) or orthogonal to it ($\theta/2=\phi_\beta + \pi/2$). Since in this case the quadrature variance along $\beta$ is $W_1/4$, while along the orthogonal direction [$\varphi=\phi_\beta +\pi/2$] the variance is $1/4W_2$, the short axis is along $\beta$ if $W_1W_2<1$, and it is orthogonal to the direction of $\beta$ if $W_1 W_2>1$.

While Eqs.\ (\ref{eq:conv-D})--(\ref{eq:conv-theta}) show the conversion (for $|\beta|\to\infty$) from the Fock-space parameters $W_1$, $W_2$, and $K$ to the phase-space  parameters $D_0$, $b$, and $\theta$, the inverse conversion is given by equations
    \begin{align}
& W_1 = 4 [D_0 -b \cos (\theta -2\phi_\beta)],
    \label{eq:conv-W1} \\
& W_2 = \frac{D_0 -b \cos (\theta -2\phi_\beta)}{4 (D_0^2 - b^2)},
    \label{eq:conv-W2} \\
&  K = \frac{b \sin (\theta-2\phi_\beta)}{ 4 [D_0 -b \cos (\theta -2\phi_\beta)]} .
    \label{eq:conv-K}
    \end{align}

\vspace{0.3cm}

The main idea of introducing the Fock-space Gaussian state \eqref{eq:FG} is that it has a simple evolution due to resonator nonlinearity.
Let us consider the evolution {\it only} due to Hamiltonian \eqref{eq:H-r}, i.e., with $\varepsilon =\kappa =0$.
Then $\rho_{nm}(t)=\rho_{nm}(0)\exp \{-i [E_{\rm rf}(n)-E_{\rm rf}(m)]t \}$.
Comparing this phase evolution with the second line of Eq.\ (\ref{eq:FG}) and expanding the resonator frequency $\omega_{\rm r}(n)$ in Eq.\ (\ref{eq:H-r}) up to first order around $n\approx |\beta ^2|$ (assuming that nonlinearity is practically constant within the range $|n-|\beta|^2|\alt \sqrt{W_1}\,|\beta|$), we find evolution equations
    \begin{align}
& 	\dot{\phi_\beta} = -[ \omega_{\rm r}(|\beta|^2)-\omega_{\rm d}],
    \label{eq:phi-dot}\\
& \dot{K} = \frac{1}{2}\,|\beta|^2\, \frac{d\omega_{\rm r}(n)}{dn}\biggr\rvert_{|\beta|^2},
    \label{eq:K-dot}\end{align}
where we neglected discreteness of $\omega_{\rm r}(n)$.
We see that $\beta$ rotates due to detuning of the resonator frequency $\omega_{\rm r}(|\beta|^2)$ at the state center from the rotating-frame frequency $\omega_{\rm d}$ (as should be expected), while nonlinearity changes $K$, leading to accumulation of the quadratic phase factor in Eq.\,\eqref{eq:FG}.

We emphasize that a weak nonlinearity approximately preserves the Fock-space Gaussian form \eqref{eq:FG}, and therefore approximately preserves the Gaussian-state form in the phase space, assuming a large photon number $|\beta|^2$.
Since the evolution due to the drive and damping also preserves the Gaussian-state form, as discussed in Sec.\ \ref{sec:Gaussian} (for weak nonlinearity we can use approximately the same matrix elements of operator $a$ in the Fock space as for a linear oscillator), the state remains approximately Gaussian in both phase and Fock spaces during the combined evolution.

\subsection{Hybrid phase-Fock-space evolution equations}\label{subsec:evol}

We have separately described the evolution due to nonlinearity, Eqs.\ (\ref{eq:phi-dot})--(\ref{eq:K-dot}), and due to drive and damping, Eqs.\ (\ref{eq:evol-beta})--(\ref{eq:evol-theta}).
The combined evolution is simply the sum of the corresponding terms.
However, Eqs.\ (\ref{eq:evol-beta})--(\ref{eq:evol-theta}) assume the phase-space representation of Fig.\ \ref{fig:quad}, while Eq.\ (\ref{eq:K-dot}) is based on the Fock-state representation (\ref{eq:FG}).
Thus, we need to convert the equations into a common representation using the conversion formulas (\ref{eq:conv-beta})--(\ref{eq:conv-theta}).

We will characterize the evolving state by four parameters: $\beta(t)$, $W_1(t)$, $W_2(t)$, and $K(t)$.
We call it a hybrid representation, since $\beta$ is a phase-space parameter, while $W_1$, $W_2$, and $K$ originate from the Fock-space description.

As discussed in Sec.\ \ref{subsec:FG}, evolution due to nonlinearity produces Eq.\ (\ref{eq:K-dot}) for $\dot{K}$, the center $\beta$ evolves as
    \begin{equation}\label{eq:beta-dot-nodrive}
	\dot{\beta} = -i[\omega_{\rm r}(|\beta|^2)-\omega_{\rm d}]\, \beta,
    \end{equation}
while $W_1$ and $W_2$ do not evolve, $\dot{W}_1= \dot{W}_2 =0$.
Note that Eq.\ (\ref{eq:beta-dot-nodrive}) essentially implies that the average number of photons in the resonator is $\bar{n} \approx |\beta|^2$, neglecting corrections in Eq.\ (\ref{eq:nbar}).

To find evolution of parameters $W_1$, $W_2$, and $K$ due to drive and damping, we write Eqs.\ \eqref{eq:evol-D0}--\eqref{eq:evol-theta} expressing the time derivatives $\dot{D_0}$, $\dot{b}$, and $\dot{\theta}$ via the  partial derivatives over the parameters of the conversion equations \eqref{eq:conv-D}--\eqref{eq:conv-theta},
    \begin{align}
& \frac{\partial D_0}{\partial W_1}\, \dot{W_1}
+\frac{\partial D_0} {\partial W_2} \, \dot{W_2}
+\frac{\partial D_0}{\partial K}\, \dot{K}
= -\kappa D_0
    \nonumber \\
& \hspace{3.6cm} + (\kappa /4) \, \coth (\omega_{\rm r0}/2T_{\rm b}) ,
    \label{partial-sum-D0}\\
&\frac{\partial b}{\partial W_1}\, \dot{W_1}
+\frac{\partial b}{\partial W_2}\, \dot{W_2}
+\frac{\partial b}{\partial K}\dot{K} = -\kappa b,
    \\
&\frac{\partial \theta}{\partial W_1}\, \dot{W_1}
+\frac{\partial \theta}{\partial W_2}\, \dot{W_2}
+\frac{\partial \theta}{\partial K}\, \dot{K}
+2 \frac{d[{\rm arg}(\beta )]}{dt} = 0,
    \label{partial-sum-theta}
    \end{align}
where in the last term of Eq.\ (\ref{partial-sum-theta}) we need to use $\dot{\beta}=-\beta\kappa/2 -i\varepsilon$, not including the evolution (\ref{eq:beta-dot-nodrive}) due to detuning.
This is because the evolution (\ref{eq:beta-dot-nodrive}) compensates the right-hand-side term of Eq.\ (\ref{eq:evol-theta}), which we therefore do not write in Eq.\ (\ref{partial-sum-theta}).
Another justification of writing Eq.\ (\ref{partial-sum-theta}) in this way is that we consider evolution only due to drive and damping [not due to detuning, which is already considered in Eq.\ (\ref{eq:beta-dot-nodrive})]; then the angle $\theta$ does not change in time, and we should exclude the detuning term from $\dot{\beta}$.

Equations (\ref{partial-sum-D0})--(\ref{partial-sum-theta}) with the partial derivatives obtained from Eqs.\ \eqref{eq:conv-D}--\eqref{eq:conv-theta}, give us a system of three linear equations for $\dot{W}_1$, $\dot{W}_2$, and $\dot{K}$.
Solving this system, we find
    \begin{align}
& \dot{W_1} = 8KW_1\, \text{Re} (\varepsilon /\beta ) + \kappa \, [\coth (\omega_{\rm r0}/2T_{\rm b})  - W_1] ,
    \label{eq:evol-drive-W1}\\
& \dot{W_2} = 8KW_2 \, \text{Re} ( \varepsilon /\beta )
    \nonumber \\
&\hspace{0.9cm} + \kappa W_2 [1-W_2(1+16K^2)\coth (\omega_{\rm r0}/2T_{\rm b})] ,
    \\
&  \dot{K} = \frac{1}{4} [ (W_1W_2)^{-1} - (1+16K^2)] \,
\text{Re} (\varepsilon /\beta )
    \nonumber \\
&\hspace{0.9cm}     -\kappa\, (K/W_1) \coth (\omega_{\rm r0}/2T_{\rm b}).
    \label{eq:evol-drive-K}
    \end{align}
Note that in the term $\coth (\omega_{\rm r0}/2T_{\rm b})$ we neglect changing resonator frequency because of the weak nonlinearity assumption.
In the special case when $\kappa=0$, Eqs.\,\eqref{eq:evol-drive-W1}--\eqref{eq:evol-drive-K} reduce to  Eq.\,(47) of Ref.\,\cite{Khezri2016}.

Finally, combining the terms from Eqs.\ (\ref{eq:K-dot})--(\ref{eq:beta-dot-nodrive}) (for evolution due to nonlinearity) and from Eqs.\,\eqref{eq:evol-drive-W1}--\eqref{eq:evol-drive-K} (for evolution due to drive and damping), we obtain the hybrid phase-Fock-space evolution equations
    \begin{align}
& \dot{\beta} = -i [ \omega_{\rm r}(\bar{n})-\omega_{\rm d}]\, \beta -\frac{\kappa}{2}\,\beta -i\varepsilon, \,\,\,\, \bar{n} \approx |\beta|^2,
    \label{eq:hybrid-beta} \\
& \dot{W}_1 = 8KW_1 \,\text{Re} ( \varepsilon /\beta )
+ \kappa \, [\coth (\omega_{\rm r0}/2T_{\rm b}) - W_1] ,
     \label{eq:hybrid-W1} \\
& \dot{W}_2 = 8KW_2 \, \text{Re} ( \varepsilon /\beta )
    \nonumber \\
&\hspace{0.9cm}     + \kappa W_2 [1- W_2(1+16K^2)\coth (\omega_{\rm r0}/2T_{\rm b})] ,
     \label{eq:hybrid-W2} \\
& \dot{K} = \left( \frac{1}{4W_1W_2}-\frac{1+16K^2}{4}\right)
\text{Re} (\varepsilon /\beta )
     \nonumber \\
& \hspace{0.7cm}  - \frac{\kappa K}{W_1} \coth (\omega_{\rm r0}/2T_{\rm b}) + \frac{1}{2} \, |\beta|^2 \, \frac{d\omega_{\rm r}(n)}{dn}\biggr\rvert_{n=|\beta|^2}. \,\, \label{eq:hybrid-K}
    \end{align}
Evolution equations \eqref{eq:hybrid-beta}--\eqref{eq:hybrid-K} complemented with the conversion formulas  \eqref{eq:conv-D}--\eqref{eq:conv-theta} are the \emph{main result} of this paper.
To our knowledge, this approach to the quantum evolution of a weakly nonlinear resonator has never been used previously.

Equations \eqref{eq:hybrid-beta}--\eqref{eq:hybrid-K} describe evolution of five real parameters of a Gaussian state.
Equation \eqref{eq:hybrid-beta} describing evolution of the state center (2 real parameters) is decoupled from the other three equations.
The equations are approximate and assume $|\beta|\gg 1$ (more detailed discussion later); in general an evolving nonlinear resonator cannot be described by a Gaussian state exactly.
In spite of the requirement $|\beta|\gg 1$, Eqs.\ \eqref{eq:hybrid-beta}--\eqref{eq:hybrid-K} can be used to numerically analyze evolution starting even from $\beta =0$ with a good accuracy (the numerical results are discussed later).
There is no divergence of $\text{Re}(\varepsilon/\beta)$ in Eqs.\ \eqref{eq:hybrid-W1}--\eqref{eq:hybrid-K} at $\beta =0$ because if $\beta (t_0)=0$, then close to this time moment $\beta=-i\varepsilon (t-t_0)$ and therefore $\text{Re}(\varepsilon/\beta)=\text{Re}[i/(t-t_0)]=0$.
A numerical divergence can be easily avoided by shifting the denominator of $\text{Re}(\varepsilon/\beta)$ by a negligible amount.

Equation \eqref{eq:hybrid-beta} has a simple physical meaning; it takes into account that the resonator frequency $\omega_{\rm r} (n)$ changes with the photon number $n$, and approximates $n$ with the average photon number $\bar{n}\approx |\beta|^2$.
One may think that a simple generalization of Eq.\ \eqref{eq:hybrid-beta} is to use a more accurate value for $\bar{n}$ from  Eq.\,\eqref{eq:nbar} in $\omega_{\rm r} (\bar{n})$ [it would also require conversion equations \eqref{eq:nth-r} and \eqref{eq:conv-D}--\eqref{eq:conv-theta}].
However, numerical simulations show that this correction does not always give a better agreement with full master equation simulations using Eq.\ (\ref{eq:master}). Because of that, we do not use this correction in the numerical analysis in  Secs.\,\ref{sec:numerical} and \ref{sec:3dB}.

Note that Eqs.\ \eqref{eq:hybrid-beta}--\eqref{eq:hybrid-K} permit three natural rescalings.
First, by rescaling the time axis, it is possible to use $\kappa =1$.
Second, since discreteness of $n$ is not important in our approach, we can rescale the $n$ axis and normalize nonlinearity, for example setting $d\omega_{\rm r}(n)/dn |_{n=0} =\pm 1$.
Third, non-zero bath temperature $T_{\rm b}$ is equivalent to
rescaling $W_1\to W_1 \coth (\omega_{\rm r0}/2T_{\rm b})$ and $W_2\to W_2/\coth (\omega_{\rm r0}/2T_{\rm b})$, while using $T_{\rm b}=0$ in Eqs.\ \eqref{eq:hybrid-beta}--\eqref{eq:hybrid-K}; this leads to $D_0 \to D_0 \coth (\omega_{\rm r0}/2T_{\rm b})$ and $b\to b \coth (\omega_{\rm r0}/2T_{\rm b})$, with unchanged $\beta$ and $\theta$.

Equations \eqref{eq:hybrid-W1}--\eqref{eq:hybrid-K} describe evolution of the Fock-space parameters $W_1$, $W_2$, and $K$. It is also possible to write evolution equations for the phase-space parameters $D_0$, $b$, and $\theta$. Note that without the last term in Eq.\ (\ref{eq:hybrid-K}), Eqs.\ (\ref{eq:hybrid-W1})--(\ref{eq:hybrid-K}) exactly correspond to Eqs.\ (\ref{eq:evol-D0})--(\ref{eq:evol-theta}). Therefore, we only need to convert the last term in (\ref{eq:hybrid-K}) into the phase space, that can be done by using partial derivatives from the conversion relations (\ref{eq:conv-W1})--(\ref{eq:conv-K}). In this way we obtain the following evolution equations,
    \begin{align}
& \dot{D}_0 = -\kappa D_0 + (\kappa/4) \coth (\omega_{\rm r0}/2T_{\rm b}) +2\eta_\beta |\beta|^2 b \sin (\Delta\theta) ,
     \label{eq:hybrid-D0} \\
& \dot{b} = -\kappa b +2\eta_\beta |\beta |^2 D_0 \sin (\Delta\theta) ,
     \label{eq:hybrid-b} \\
& \frac{d(\Delta\theta)}{dt} = 2\, \text{Re} (\varepsilon /\beta ) - 2\eta_\beta |\beta |^2 \, \frac{b-D_0\cos (\Delta\theta)}{b} ,
\label{eq:hybrid-theta}
    \end{align}
where $\Delta \theta \equiv \theta -2\,{\rm arg} (\beta)$,   $\eta_\beta \equiv d\omega_{\rm r} (n)/dn\big|_{n=|\beta|^2}$, and evolution of $\beta$ is still given by Eq.\ (\ref{eq:hybrid-beta}). Note that divergence in Eq.\ (\ref{eq:hybrid-theta}) at $\beta =0$ can be avoided numerically in the same way as discussed above: by a negligible shift of $\beta$. The divergence in  Eq.\ (\ref{eq:hybrid-theta}) at $b=0$ can also be avoided numerically by a negligible increase of $b$ (physically, this divergence is because $\Delta\theta$ is undefined at $b=0$).
Equations (\ref{eq:hybrid-D0})--(\ref{eq:hybrid-theta}) are equivalent to Eqs.\ (\ref{eq:hybrid-W1})--(\ref{eq:hybrid-K}). We have checked this equivalence numerically. However, in the simulations discussed below we used Eqs.\ (\ref{eq:hybrid-W1})--(\ref{eq:hybrid-K}) rather than Eqs.\  (\ref{eq:hybrid-D0})--(\ref{eq:hybrid-theta}). One of the reasons for our preference is that evolution of $W_1$, $W_2$, and $K$ is always smooth, while $\Delta \theta$ evolves very fast when $b$ approaches zero, thus potentially creating a problem with numerical solution of differential equations (even though our simulations never suffered from this potential problem).

Note that from Eqs.\ (\ref{eq:hybrid-D0}) and (\ref{eq:hybrid-b}) we can obtain
    \begin{align}
& \frac{d}{dt}(D_0\pm b) =- [\kappa \mp  2\eta_\beta |\beta|^2 \sin (\Delta\theta) ] \, (D_0\pm b)
    \nonumber \\
&\hspace{2.1cm} + (\kappa/4) \coth (\omega_{\rm r0}/2T_{\rm b}) ,
    \label{eq:evol-Dpmb}\end{align}
which shows that for the maximum-variance and minimum-variance quadratures, the effective damping rate is different, $\kappa_{\rm eff}=\kappa\mp 2\eta_\beta |\beta|^2 \sin (\Delta\theta)$, and changes with time. Similarly, the effective bath temperature is also different, $\coth (\omega_{\rm r0}/2T_{\rm b}) \to  (\kappa/ \kappa_{\rm eff}) \coth (\omega_{\rm r0}/2T_{\rm b})$. Discussion in terms of different effective damping rates for the two quadratures makes an obvious connection to the case of a parametric drive with doubled frequency.

We have checked that Eqs.\ (\ref{eq:hybrid-D0})--(\ref{eq:hybrid-theta}) are consistent with the results of Ref.\ \cite{Ludwig1975} for Gaussian variances of classical fluctuations around the trajectory (\ref{eq:hybrid-beta}), caused by classical (complex) white noise  $\sqrt{\kappa}\, \zeta (t)$ applied to the resonator, with the correlation function $\langle \zeta^*(t)\, \zeta(t')\rangle =(1/2) \coth (\omega_{\rm r}/2T_{\rm b})\, \delta(t-t')$, $\langle \zeta(t)\, \zeta(t')\rangle =0$ (as in, e.g., \cite{Korotkov2016}). Note, however, that in order to get correct equations, we had to exchange $B$ with $B^\dagger$ in Eq. (3.2.4) of Ref.\ \cite{Ludwig1975}. The correspondence between Eqs.\ (\ref{eq:hybrid-D0})--(\ref{eq:hybrid-theta}) and results of Ref.\ \cite{Ludwig1975} confirms that the quantum squeezing is similar to squeezing of classical fluctuations, and it also shows that our approach is physically similar to linearization of fluctuations around the classical trajectory within the Gaussian approximation.

In Appendix \ref{app:steady} we derive analytical results for $D_0$, $b$, and $\Delta \theta$ in the steady state and discuss their equivalence to the results of Refs.\ \cite{Drummond1980} and \cite{Dykman-2012} for a Duffing oscillator (Kerr nonlinearity).

\section{Numerical accuracy}\label{sec:numerical}

In this section we discuss numerical accuracy of our approach.
We start with analyzing fidelity of the conversion between the Gaussian and Fock-space Gaussian states, and then discuss numerical accuracy of the hybrid phase-Fock-space evolution equations by comparing results with full simulation.

\subsection{Fidelity of the conversion} \label{subsec:conversion-fidelity}

As was discussed in Sec.\ \ref{subsec:FG}, the Gaussian state (\ref{eq:W-norm}) is approximately equal to the Fock-space Gaussian state (\ref{eq:FG}) with the conversion relations (\ref{eq:conv-beta})--(\ref{eq:conv-theta}), in the case of large photon numbers, $|\beta|^2 \gg 1$.
Let us check the accuracy of this conversion numerically.
For that we calculate the overlap fidelity $F$ between the states (\ref{eq:W-norm}) and (\ref{eq:FG}) using the standard definition \cite{Nielsen2000}
    \begin{equation}
F =\frac{\Big( \text{Tr} \sqrt{ \sqrt{\rho_1} \,\rho_2 \sqrt{\rho_1}} \, \Big)^2}{{\rm Tr} (\rho_1) \, {\rm Tr} (\rho_2)} \, ,
    \label{fidelity-def}\end{equation}
where $\rho_1$ and $\rho_2$ are the density matrices of the compared states.
Note that for normalized states the denominator in Eq.\ (\ref{fidelity-def}) is not needed, but we use the more general version (\ref{fidelity-def}) because the Fock-space Gaussian state (\ref{eq:FG}) is not exactly normalized.
When at least one of the states is pure, Eq.\ (\ref{fidelity-def}) reduces to the usual state overlap, e.g., $F=\langle \psi_1 | \rho_2 |\psi_1\rangle={\rm Tr}(\rho_1\rho_2)$ if $\rho_1=|\psi_1\rangle\langle \psi_1|$ and both states are normalized.

To find the conversion fidelity for a Gaussian state with parameters $\beta$, $D_0$, $b$, and $\theta$, we use conversion relations (\ref{eq:conv-D})--(\ref{eq:conv-theta}) to find corresponding parameters $W_1$, $W_2$, and $K$ [$\beta$ is the same unless we use the correction (\ref{eq:a-aver-FG})], which gives us the Fock-space Gaussian state (\ref{eq:FG}).
Then we calculate exact Fock-space representation of the Gaussian state of \eqref{eq:W-norm} using Eq.\ (\ref{eq:DSTS}) with parameters $|\xi|$ and $n_{\rm th}$ obtained from the relations (\ref{eq:nth-r}) (using $\alpha=\beta$ and $\Theta=\theta$).
Finally, we use Eq.\ (\ref{fidelity-def}) in the Fock space to find the fidelity $F$ between the Gaussian and Fock-space Gaussian states.
Note that $F$ does not depend on the phase ${\rm arg}(\beta )$ for a fixed value of $\theta/2-{\rm arg}(\beta)$, so it is sufficient to consider ${\rm arg}(\beta )=0$, i.e.,  $\beta =|\beta |$; this is what we assume below in the numerical analysis of the conversion fidelity; in this case $\theta/2-{\rm arg}(\beta)\to \theta/2$.

\begin{figure}[t]
	\begin{center}
\includegraphics[width=0.98\columnwidth]{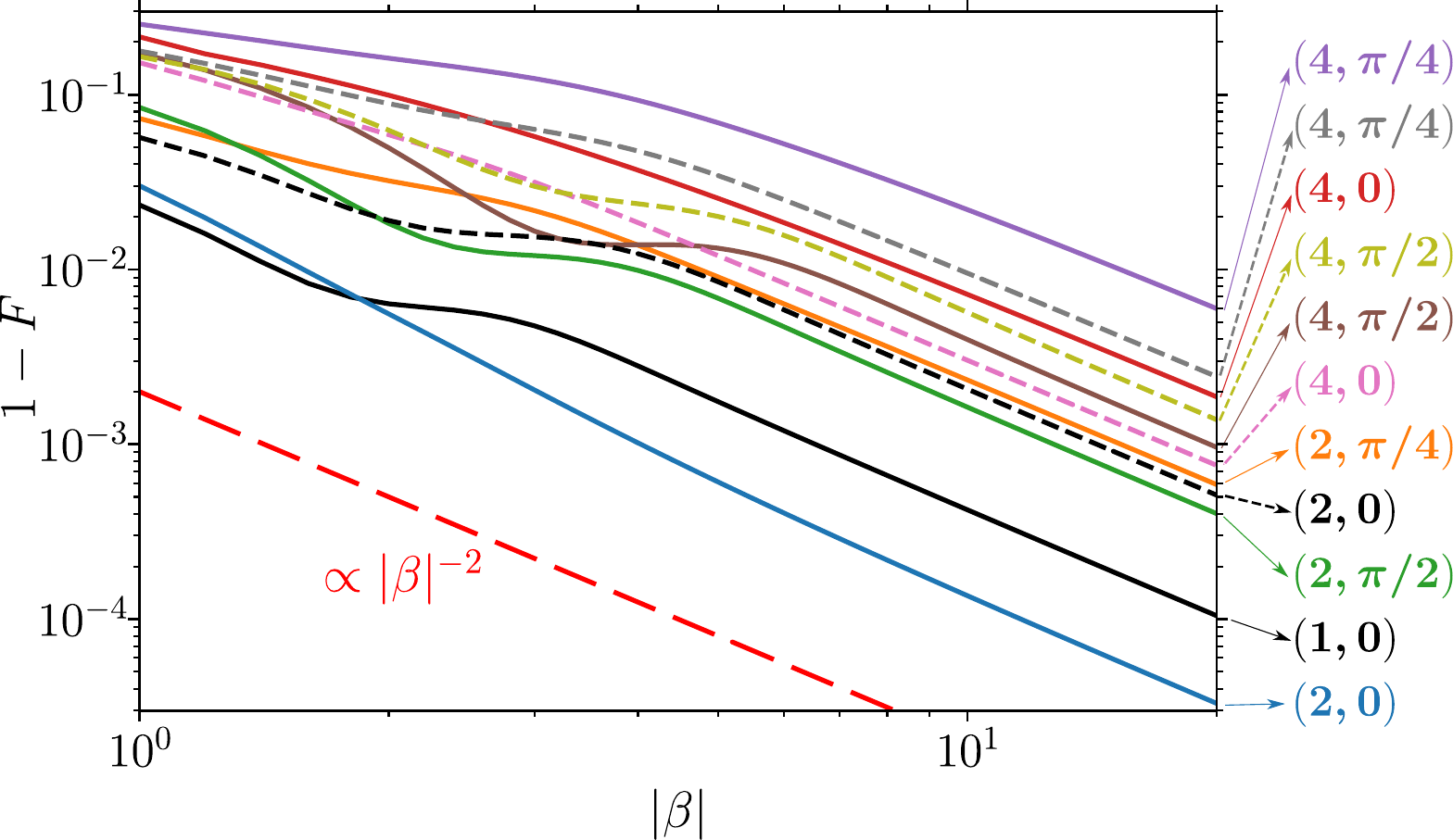}
    \end{center}
    \caption{
Infidelity $1-F$ between the Gaussian and Fock-space Gaussian states as a function of (real)  $\beta$ for several values of the parameters $4(D_0+b)$ and $\theta/2$ (labeled respectively at the right side), for  $n_{\rm th}=0$ (solid lines) and  $n_{\rm th}=1/2$ (dashed lines). At large $|\beta|$ all lines show the scaling  $|\beta|^{-2}$, illustrated by the long-dashed line.
    }\label{fig:conv-beta}
\end{figure}

Figure \ref{fig:conv-beta} shows infidelity $1-F$ as a function of $|\beta|$ on a log-log scale for several values of other parameters: $4(D_0+b)=1$, 2, and 4 (this parameter is the long-axis variance compared with the coherent state; we call it ``unsqueezing factor''), $\theta /2 =0$, $\pi/2$, and $\pi/4$ (this is the direction of the short axis in Fig.\ \ref{fig:quad}), $n_{\rm th}=0$ and $1/2$.
The lines in Fig.\ \ref{fig:conv-beta} are labeled with a pair of numbers: $4(D_0+b)$  and $\theta/2$; solid and dashed lines correspond to $n_{\rm th}=0$ and $1/2$ respectively. Note that there is no dependence on $\theta$ when $4(D_0+b)=1+2n_{\rm th}$ [see Eq.\ (\ref{eq:nth-r})], then we show only the line $\theta=0$;  also note that for $n_{\rm th}=1/2$ it is always $4(D_0+b)\geq 2$.

Most importantly, we see that all lines in Fig.\ \ref{fig:conv-beta} show the scaling $1-F\propto |\beta|^{-2}$ at large $|\beta|$ (this scaling is illustrated by the long-dashed line). The deviation from this dependence at small $|\beta|$ is mainly caused by two reasons. First, the ``shoulder'' feature may develop when
$|\beta | < 3\sqrt{4(D_0-b\cos \theta)}\leq 3 \sqrt{4(D_0+b)}$ because then $|\beta | < 3\sqrt{W_1}$ in Eq.\ (\ref{eq:FG}) and thus the Gaussian approximation near $n=0$ becomes inaccurate (less than 3 standard deviations). Second, deviation from the scaling $|\beta|^{-2}$ starts to develop when $1 -F \agt 0.05$ because $F$ cannot exceed 1; actually, a natural metric for distance between the states is ${\rm arccos}(\sqrt{F})$ \cite{Nielsen2000}, which is approximately $\sqrt{1-F}$ when  $1-F\ll 1$; for this metric the above condition is $\sqrt{1-F}\agt 0.22$. From Fig.\ \ref{fig:conv-beta} we conclude that the scaling $1-F\propto |\beta|^{-2}$ is almost perfect if $|\beta |>3\sqrt{4(D_0+b)}$ and $1-F < 0.05$.

\begin{figure}[t]
	\begin{center}
\includegraphics[width=1\columnwidth]{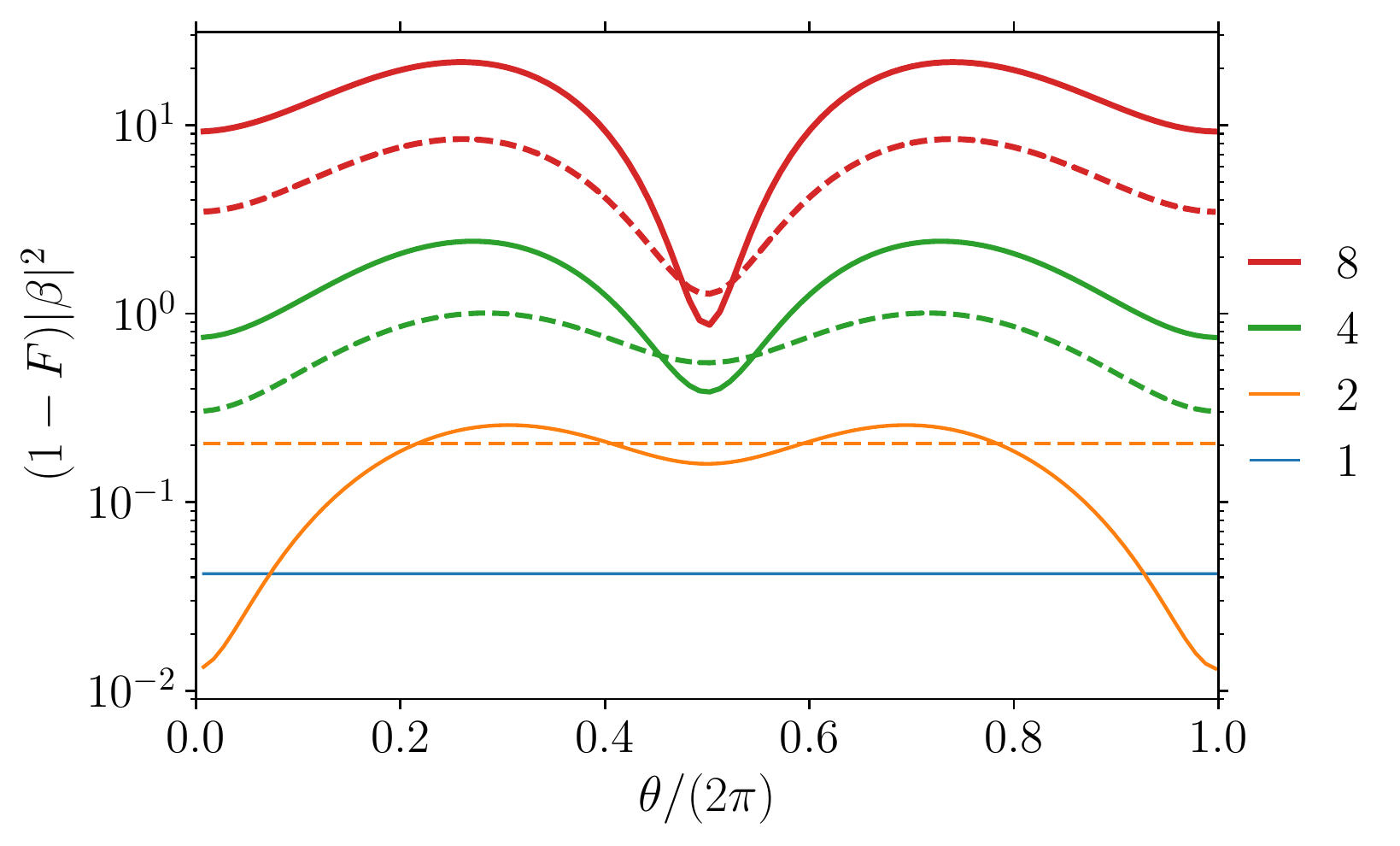}
    \end{center}
    \caption{
Scaled infidelity $(1-F)|\beta |^2$ as a function of the short-axis angle $\theta/2$. Solid lines are for $n_{\rm th}=0$ (pure states) and $4(D_0+b)=8$, 4, 2, and 1 (top to bottom); dashed lines are for $n_{\rm th}=1/2$ and $4(D_0+b)=8$, 4, and 2 (top to bottom). We used $\beta=40$, which is sufficiently large so that the presented results do not depend on $|\beta|$.
    }\label{fig:conv-theta}
\end{figure}

Figure \ref{fig:conv-theta} shows the scaled infidelity $(1-F)|\beta|^2$ for sufficiently large $|\beta|$ (here we used $\beta =40$), as a function of the short-axis angle $\theta/2$. We used parameters $4(D_0+b)=1$, 2, 4, and 8, while $n_{\rm th}=0$ (solid lines) and $1/2$ (dashed lines).
As expected, we see no dependence on $\theta/2$ when $4(D_0+b)=1+2n_{\rm th}$, since in this case the long-axis and short-axis variances coincide, $D_0+b=D_0-b$. When $4(D_0+b)>1+2n_{\rm th}$, the local minima of the infidelity are reached at $\theta/2=0$ and $\theta/2=\pi/2$; both these cases correspond to $K=0$ in Eq.\ \eqref{eq:FG} [note that $K=0$ minimizes the state center shift in Eq.\ (\ref{eq:a-aver-FG}), which affects infidelity, as discussed below]. For relatively small values of $4(D_0+b)$, the minimum is reached at $\theta/2=0$ (``photon number squeezing''), while at larger $4(D_0+b)$, the minimum infidelity is at $\theta/2=\pi/2$ (``phase squeezing'').
The maximum infidelity is reached when $\theta/2$ is (crudely) near $\pm \pi/4$. Note that the infidelity dependence on $\theta/2$ has a period of $\pi$, and the dependence is symmetric about the points $\theta/2=0$ and  $\theta/2=\pi/2$.

\begin{figure}[t]
	\begin{center}
\includegraphics[width=.95\columnwidth]{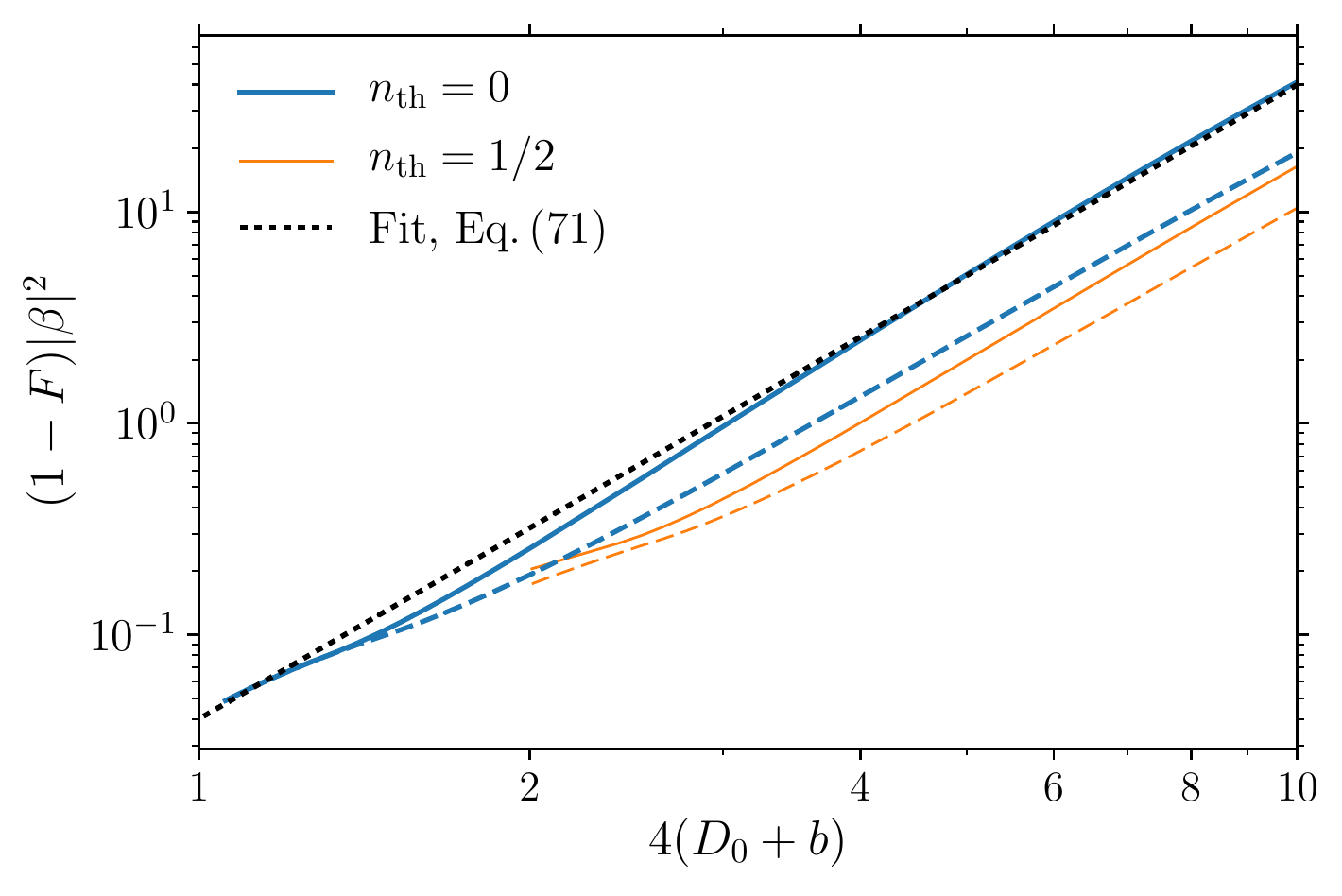}
    \end{center}
    \caption{
Solid lines: scaled infidelity $(1-F)|\beta |^2$ maximized over the angle $\theta/2$ (for $\beta =40$), as a function of the quadrupled long-axis variance $4(D_0+b)$. The upper (blue) solid line is for $n_{\rm th}=0$, the lower (orange) solid line is for $n_{\rm th}=1/2$.
For the corresponding dashed lines we used the correction to the state center via Eq.\ \eqref{eq:a-aver-FG}. The black dotted line is a crude fit given by Eq.\ (\ref{1-F-num}).
    }\label{fig:conv-long-axis}
\end{figure}

The upper (blue) solid line in Fig.\ \ref{fig:conv-long-axis} shows the scaled infidelity $(1-F)|\beta|^2$ maximized over the angle $\theta/2$ (the worst case), as a function of the unsqueezing factor $4(D_0+b)$ (long-axis variance in units of the coherent state variance) for the case $n_{\rm th}=0$ (zero effective temperature). We see that this line can be approximately fitted by the formula
    \begin{equation}\label{1-F-num}
	1-F \approx 0.04 \, \frac{[4(D_0+b)]^3}{|\beta|^2},
    \end{equation}
which is drawn as the dotted black line.

The infidelity scaling $1-F\propto (D_0+b)^3$ can be crudely understood as a consequence of the Fock-space Gaussian state center shift described by Eq.\ (\ref{eq:a-aver-FG}). Considering for simplicity the case $K=0$ and $W_1=W_2\ll 1$ (i.e., $n_{\rm th}=0$,  $\theta/2=0$ -- see Figs.\ \ref{fig:conv-beta} and $\ref{fig:conv-theta}$), we find that the state center is shifted by $\Delta |\beta| \approx -(8W_2|\beta|)^{-1}\approx - (D_0+b)/(2|\beta|)$ along the short axis.
The  relative shift compared with the ``width'' of the state along the short axis is then $\Delta |\beta|/ \sqrt{D_0-b}\approx -[4(D_0+b)]^{3/2}/(4 |\beta|)$.
Since the infidelity scales quadratically with this relative shift, $1-F \propto (\Delta |\beta|/ \sqrt{D_0-b})^2$, we obtain the scaling  $1-F \propto [4(D_0+b)]^3/|\beta|^2$.

The same numerical scaling of the infidelity in Eq.\ (\ref{1-F-num}) indicates that the state center shift may play a significant role in fidelity reduction. To check this hypothesis, we used the correction from Eq.\ (\ref{eq:a-aver-FG}) to produce Gaussian and Fock-space Gaussian states with the same $\langle a\rangle$ by making a small compensating shift of $\beta$.
The corresponding result for the infidelity $1-F$ is shown by the upper (blue) dashed line in Fig.\ \ref{fig:conv-long-axis}.
As we see, the correction has really decreased the infidelity;
however, the improvement is only by a factor of about 2, so the scaling is approximately the same as in Eq.\ (\ref{1-F-num}), with the factor 0.04 replaced by 0.02.
We have also checked that numerical optimization of the infidelity over the center shift of the Fock-space Gaussian state [instead of using Eq.\ (\ref{eq:a-aver-FG})] produces practically the same result.
The infidelity decrease by a factor of about 2 can be crudely understood in the following way.
The Fock-space Gaussian state has a slightly crescent (non-elliptical) shape of the Wigner function in the phase plane.
Slightly shifting its center, it is possible to improve the state fidelity compared with the Gaussian state (which has a perfect elliptical shape); however, this improvement cannot be very significant.

Now let us discuss the lower (orange) lines in Fig.\ \ref{fig:conv-long-axis}, for which $n_{\rm th}=1/2$ (i.e., effective temperature is $T_{\rm eff}=0.91\, \omega_{\rm r}$); as above, the dashed line takes into account the center  correction (\ref{eq:a-aver-FG}), while the solid line is without the correction.
We see that non-zero $n_{\rm th}$ improves the fidelity compared with the case $n_{\rm th}=0$ for the same long-axis variance $D_0+b$ (the short-axis variance in this case is increased by a factor of 4).
The improvement can be qualitatively understood using the above derivation based on the state center shift: since the short-axis ``width'' is now larger, the relative inaccuracy is smaller, thus decreasing the infidelity. Note, however, that such derivation would predict infidelity reduction by a factor of 4, while numerically the distance between the upper and lower solid lines in Fig.\ \ref{fig:conv-long-axis} is less than a factor of 2.5.
Comparing the solid and dashed orange lines, we see that the state center correction decreases the infidelity; however, the improvement is only by crudely a factor of 1.5, even less than in the zero-temperature case.

We can make the following conclusions from the numerical results discussed in this section.
First, the infidelity of the conversion between the Gaussian and Fock-space Gaussian states is not larger than in Eq.\ (\ref{1-F-num}), so the conversion becomes almost perfect for sufficiently large $|\beta|$.
Second, correction (\ref{eq:a-aver-FG}) to the state center improves the fidelity; however, the improvement is not very significant (we will not use this correction in analyzing the evolution). Let us also note that the change of effective temperature from zero to $0.9\, \omega_{\rm r}$ ($n_{\rm th}=1/2$) did not produce a very significant change in the infidelity.

\subsection{Accuracy of the hybrid phase-Fock-space evolution equations}\label{subsec:fid-hybrid}

The main result of this paper is the hybrid phase-Fock-space evolution equations \eqref{eq:hybrid-beta}--\eqref{eq:hybrid-K}, which permit a very efficient approximate simulation of the state dynamics for a slightly nonlinear resonator in the large-photon-number regime.
In contrast, full simulation using the master equation (\ref{eq:master}) is highly resource-consuming in this regime because of large Hilbert space.
In this section we numerically analyze the accuracy of our hybrid equations by comparing the results with the full master equation simulation.

For the numerical analysis let us consider a constant drive, $\varepsilon (t)=\varepsilon$, and a constant (Kerr) nonlinearity,
    \begin{equation}
	\omega_{\rm r}(n)= \omega_{\rm r0}+ n\eta,
    \label{Kerr}\end{equation}
which corresponds to the rotating-frame resonator energy levels $E_{\rm rf}(n)=(\omega_{\rm r0}-\omega_{\rm d})n +n(n-1)\eta/2$.
We also assume that initial state is vacuum, $\rho (0)=|0\rangle\langle 0|$.
Note that the hybrid evolution equations still work well when initial state is vacuum, because for sufficiently weak nonlinearity, the photon number becomes large before the effects due to nonlinearity (e.g., squeezing) become important. Also note that in RWA the considered resonator Hamiltonian is equivalent to $H_{\rm r}^{\rm lf}=P^2/(2m) + (m/2)\,\tilde{\omega}_{\rm r0}^2 X^2 +(\eta/3)\, m^2 \tilde{\omega}_{\rm r0}^2 X^4$, where $\tilde{\omega}_{\rm r0}=\omega_{\rm r0}-\eta$. The difference between the first-excitation frequency $\omega_{\rm r0}$ and the ``plasma frequency'' $\tilde{\omega}_{\rm r0}$ for a Duffing oscillator is negligible  because we focus on the regime of large $n$.

In the considered case, the RWA dynamics described by the master equation (\ref{eq:master}) depends on five parameters: nonlinearity $\eta$, drive amplitude $\varepsilon$, initial detuning $\omega_{\rm r0}-\omega_{\rm d}$, damping rate $\kappa$, and bath temperature $T_{\rm b}$ characterized by the bath photon number $n_{\rm b}$ via Eq.\ (\ref{n-b-def}).
Rescaling the time axis (using $\kappa^{-1}$ as the time unit), it is easy to see that the dynamics depends on four dimensionless parameters: $\eta/\kappa$, $\varepsilon/\kappa$, $(\omega_{\rm r0}-\omega_{\rm d})/\kappa$, and $n_{\rm b}$.

For simulation using the hybrid evolution equations \eqref{eq:hybrid-beta}--\eqref{eq:hybrid-K}, it is possible to further reduce the number of free parameters from four to only two (this is not possible for full master equation simulation).
Since discreteness of $n$ is not used in Eqs.\ \eqref{eq:hybrid-beta}--\eqref{eq:hybrid-K}, it is possible to rescale $n$-axis using $\kappa /|\eta |$ as the unit of $n$; this eliminates nonlinearity as a free parameter, $d(\omega_{\rm r}/\kappa)/d[n/(\kappa/|\eta| )]=\pm 1$ (the sign here is the sign of $\eta$).
This rescaling renormalizes the drive amplitude as $(\varepsilon /\kappa)/\sqrt{\kappa /|\eta |}$, while not affecting dimensionless detuning.
Furthermore, it is possible to rescale $W_1$ and $W_2$ using $1/\coth (\omega_{\rm r0}/2T_{\rm b})$ and $\coth (\omega_{\rm r0}/2T_{\rm b})$ respectively; this eliminates bath temperature as a free parameter, such that it can always be assumed zero.  Then the rescaled dynamics is determined by only two free parameters: $\varepsilon \sqrt{|\eta |}/\kappa^{3/2}$ and $(\omega_{\rm r0}-\omega_{\rm d})/\kappa$, and
we can use Eqs.\ \eqref{eq:hybrid-beta}--\eqref{eq:hybrid-K} with the following  parameters: $\kappa \to 1$, $d\omega_{\rm r}/dn \to \pm 1$ (depending on the sign of $\eta$), $\varepsilon \to \varepsilon \sqrt{|\eta |}/\kappa^{3/2}$, $\omega_{\rm r0}-\omega_{\rm d}\to (\omega_{\rm r0}-\omega_{\rm d})/\kappa$, and $T_{\rm b}\to 0$;
this automatically rescales $\beta$ as $\beta \to \beta /\sqrt{\kappa /|\eta |}$,  time as $t\to \kappa t$, variables $W_1$ and $W_2$ as $W_1\to W_1\coth (\omega_{\rm r0}/2T_{\rm b})$ and $W_2\to W_2/\coth (\omega_{\rm r0}/2T_{\rm b})$,  while $K$ does not change.

To check accuracy of the hybrid phase-Fock-space evolution equations,
let us calculate the time-dependent fidelity $F(t)$ [Eq.\,(\ref{fidelity-def})]  between the exact solution $\rho_{\rm m}(t)$ of the master equation (\ref{eq:master}) and the state $\rho_{\rm h}(t)$ obtained from our approximate hybrid equations \eqref{eq:hybrid-beta}--\eqref{eq:hybrid-K}.
Note that in the hybrid method we evolve variables $\beta$, $W_1$, $W_2$, and $K$, but the resulting state is always converted into a Gaussian state using Eqs.\,(\ref{eq:conv-D})--(\ref{eq:conv-theta}), so the fidelity $F(t)$ is calculated between this Gaussian state and Fock-space solution of the master equation [for that the Gaussian state is represented in the Fock space using Eq.\,(\ref{eq:DSTS})].
In simulations we will use parameters somewhat close to typical parameters in circuit QED experiments for measurement of superconducting transmon qubits; a weak nonlinearity of the resonator in this case is induced by the qubit nonlinearity; the resonator nonlinearity is much more significant when the transmon is in the ground state \cite{Khezri2016}.

\begin{figure}[t]
	\begin{center}
\includegraphics[width=.95\columnwidth]{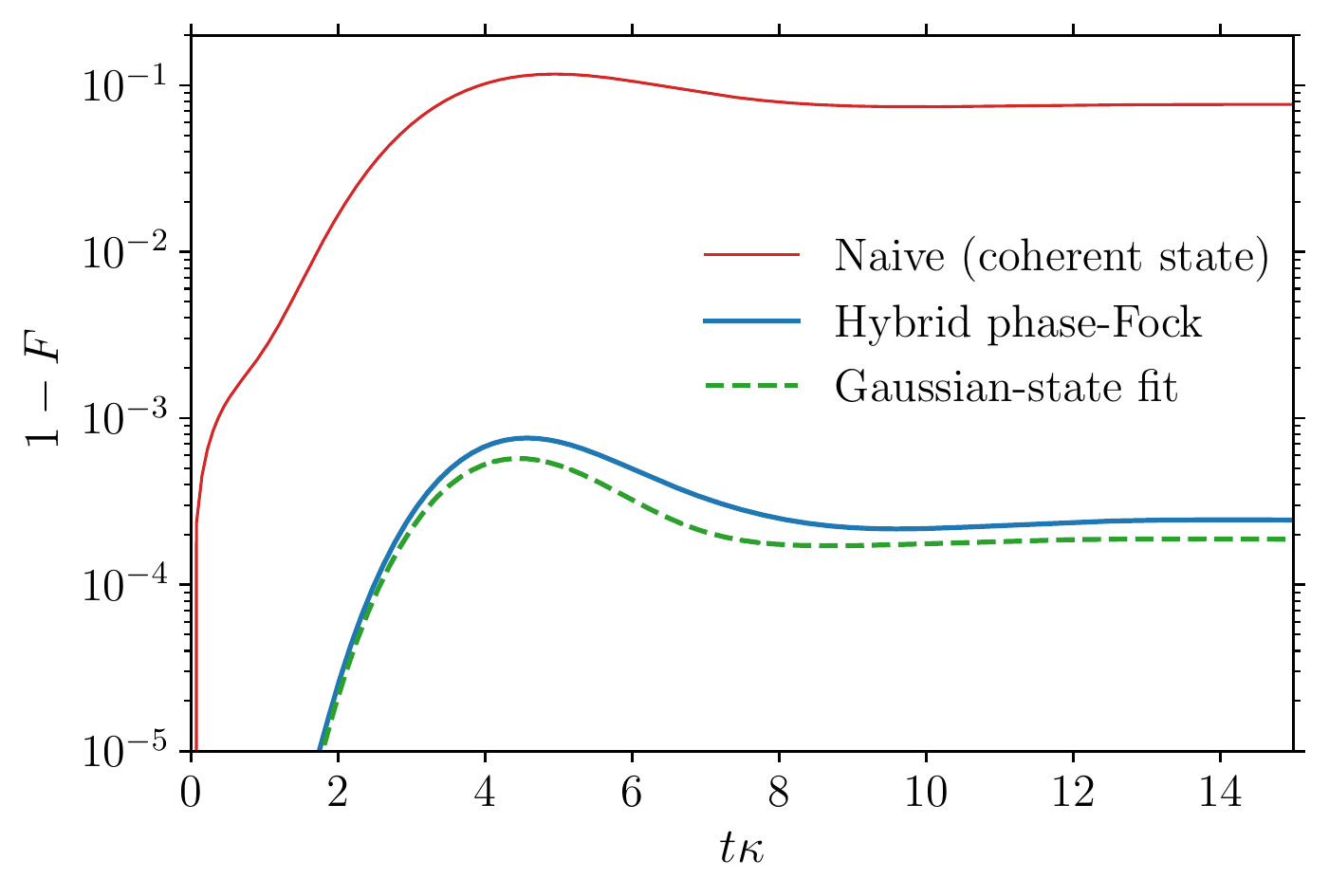}
    \end{center}
	\caption{
Blue (lower) solid line: time dependence of infidelity $1-F(t)$ between the exact solution $\rho_{\rm m}(t)$ obtained from the master equation (\ref{eq:master}) and state $\rho_{\rm h}(t)$ obtained from the hybrid evolution equations \eqref{eq:hybrid-beta}--\eqref{eq:hybrid-K}. Parameters are close to typical circuit QED parameters (see text), time $t$ is normalized by the resonator decay time $\kappa^{-1}$. Dashed green line: infidelity between $\rho_{\rm m}(t)$ and its  Gaussian-state fit. Red (upper) solid line: infidelity of the conventional approach based on coherent states.
    }\label{fig:F(t)-1}
\end{figure}

The lower (blue) solid line in Fig.\ \ref{fig:F(t)-1} shows the time-dependent infidelity $1-F$ of the calculation based on the hybrid phase-Fock-space evolution equations \eqref{eq:hybrid-beta}--\eqref{eq:hybrid-K}. Here we used parameters
$\kappa/2\pi =5\,$MHz, $\omega_{\rm r0}-\omega_{\rm d}=0$, $\eta /2\pi = -0.02\,$MHz, $\varepsilon/2\pi=32\,$MHz (this corresponds to 100 photons in the steady state), and $n_{\rm b}=3.2\times 10^{-3}$ (this corresponds to $T_{\rm b}=50\,$mK for $\omega_{\rm r0}/2\pi=6\,$GHz; we start with the vacuum state instead of the thermal state, but the difference is negligible). We see a very good accuracy provided by our approach, with infidelity below $10^{-3}$. For comparison, the upper (red) solid line shows the infidelity for the conventional naive approach, in which we assume a coherent state of the resonator, with the same center $\beta(t)$ given by Eq.\ (\ref{eq:hybrid-beta}). We see that the conventional approach fails to describe the evolution with a good accuracy, thus emphasizing importance of considering Gaussian states in our approach.

\begin{figure}[t]
	\begin{center}
\includegraphics[width=.95\columnwidth]{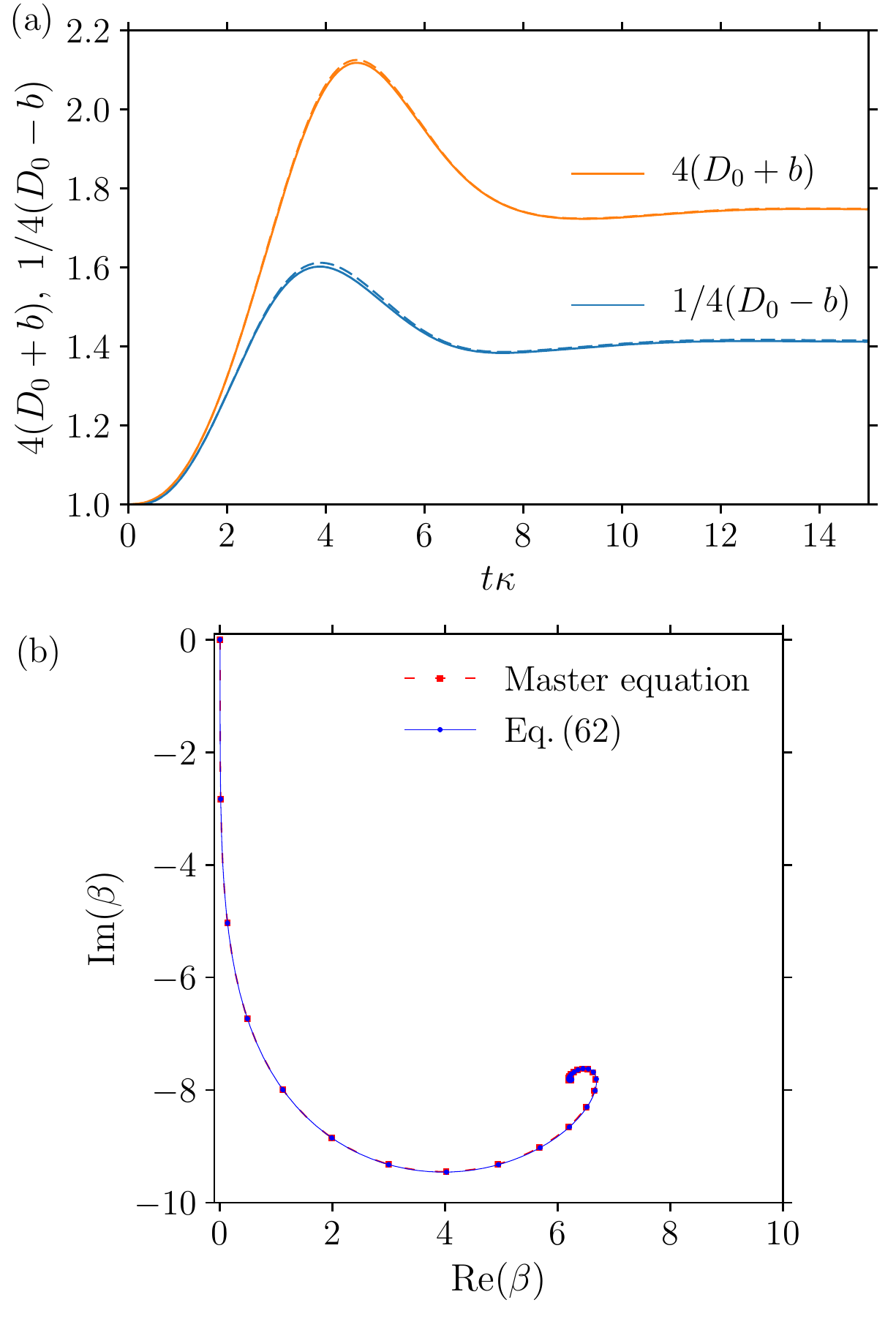}
    \end{center}
	\caption{
	Panel (a): ``Squeezing factor'' $[4(D_0-b)]^{-1}$ (lower lines) and ``unsqueezing factor'' $4(D_0+b)$ (upper lines) as functions of time, for parameters of Fig.\ \ref{fig:F(t)-1}.
	Solid lines are obtained from the hybrid evolution equations \eqref{eq:hybrid-beta}--\eqref{eq:hybrid-K}, dashed lines are obtained from the  Gaussian-state fit to the master-equation result $\rho_{\rm m}(t)$.
	Panel (b): Corresponding evolution of the state center $\beta(t)$ on the phase plane, with points spaced in time by $0.5/\kappa$.
 Solid blue line with dots is calculated using Eq.\ (\ref{eq:hybrid-beta}), almost coinciding red dashed line with squares show $\langle a \rangle$ for $\rho_{\rm m}(t)$.
	}
    \label{fig:bump}
\end{figure}

For the dashed green line in Fig.\ \ref{fig:F(t)-1}, at each time $t$
we fitted $\rho_{\rm m}(t)$ by a Gaussian state having the same values of $\langle a\rangle$, $\langle a^2\rangle$, and $\langle a^\dagger a\rangle$,
and then calculated fidelity between this Gaussian state and $\rho_{\rm m}(t)$. Therefore, the dashed line essentially shows the non-Gaussianity of the actual state  $\rho_{\rm m}(t)$ (we have checked that numerical optimization over the state center $\beta$ does not provide a noticeable further improvement of the infidelity).
Comparing the dashed green  line with the blue solid line, we see that our hybrid evolution equations \eqref{eq:hybrid-beta}--\eqref{eq:hybrid-K} describe the resonator state almost as good as this Gaussian-state fit.
We have found numerically that almost all difference between the solid blue and dashed green lines in Fig.\ \ref{fig:F(t)-1} comes from a small inaccuracy in calculation of the state center using Eq.\,\eqref{eq:hybrid-beta} [see Fig.\,\ref{fig:bump}(b)]. We tried to improve this accuracy by using $\bar{n}$ from Eq.\,\eqref{eq:nbar} for the center evolution \eqref{eq:hybrid-beta} and also by using the center correction  \eqref{eq:a-aver-FG}. While this decreased infidelity for some parameters, it increased it for some other parameters, so we decided to use the simplest equation \eqref{eq:hybrid-beta} for the state center evolution. As follows from Fig.\  \ref{fig:F(t)-1}, this already gives a very good accuracy.

To clarify the origin of the ``bump'' on the lower lines in Fig.\ \ref{fig:F(t)-1}, in Fig.\ \ref{fig:bump}(a) we show the corresponding evolution of ``squeezing parameter'' $1/[4(D_0-b)]$ (lower lines) and ``unsqueezing parameter'' $4(D_0+b)$ (upper lines). We see that the maximum infidelity in Fig.\ \ref{fig:F(t)-1} occurs at approximately the same time as the maximum unsqueezing in Fig.\ \ref{fig:bump}(a), thus hinting that the infidelity during evolution originates from a mechanism similar to the infidelity between the Gaussian and Fock-space Gaussian states estimated by Eq.\ (\ref{1-F-num}). The quantitative comparison shows that the maximum of the lower solid line in Fig.\ \ref{fig:F(t)-1} is about a factor of 4 smaller than the estimate given by Eq.\ (\ref{1-F-num}), while the steady-state infidelity is smaller than this estimate by a factor of 9.

The solid lines in Fig.\ \ref{fig:bump}(a) are calculated using the hybrid evolution equations  \eqref{eq:hybrid-beta}--\eqref{eq:hybrid-K}, while dashed lines are obtained from the Gaussian-state fit of the master-equation result $\rho_{\rm m}(t)$. We see that the dashed and solid lines are very close to each other, indicating that our hybrid approach is quite accurate in calculating the quadrature variances.

Note that for a minimum-uncertainty (pure) state, the lower and upper lines (squeezing and unsqueezing) in Fig.\ \ref{fig:bump}(a) should coincide; the ratio between these parameters is $\coth^2(\omega_{\rm r0}/2T_{\rm eff})$ -- see Eq.\ (\ref{W1/W2}). From Fig.\ \ref{fig:bump}(a) we see that the resonator state is considerably mixed, with the effective temperature $T_{\rm eff}$ significantly exceeding \cite{Dykman-2012} the bath temperature $T_{\rm b}$; for example, in the steady state $T_{\rm eff}= 98\,$mK, in contrast to $T_{\rm b}=50\,$mK. A large corresponding ratio of thermal photon numbers, $n_{\rm th}/n_{\rm b}=17.3$, indicates that the effective temperature $T_{\rm eff}$ in this case is practically independent of the bath temperature. Indeed, the same simulations with $T_{\rm b}=0$ showed a very close effective temperature, $T_{\rm eff}= 96\,$mK.

In Fig.\ \ref{fig:bump}(b) we show evolution of the state center $\beta(t)$ on the phase plane for the same parameters as in Figs.\  \ref{fig:F(t)-1} and \ref{fig:bump}(a). The dots (and squares) are separated by time intervals $0.5/\kappa$ (which is 15.9 ns); the solid blue line with dots is for calculation using Eq.\ \eqref{eq:hybrid-beta}, while the dashed red line with squares shows $\langle a \rangle$ for the master-equation simulation result $\rho_{\rm m}(t)$.
We see that Eq.\ \eqref{eq:hybrid-beta} is quite accurate for calculating the state center.  However, there is a tiny (almost unnoticeable) difference between positions of the dots and squares in Fig.\ \ref{fig:bump}(b); as mentioned above, this tiny shift is mainly responsible for the difference between the lower solid and dashed lines in Fig.\  \ref{fig:F(t)-1}. As another observation, the maximum photon number $|\beta|^2$ is achieved at almost the same time as the maximum of $4(D_0+b)$; however, we think that the infidelity bump in Fig.\  \ref{fig:F(t)-1} is caused by the maximum of $4(D_0+b)$ and not by the almost simultaneous maximum of $|\beta|^2$.

\begin{figure}[t]
	\begin{center}
\includegraphics[width=.95\columnwidth]{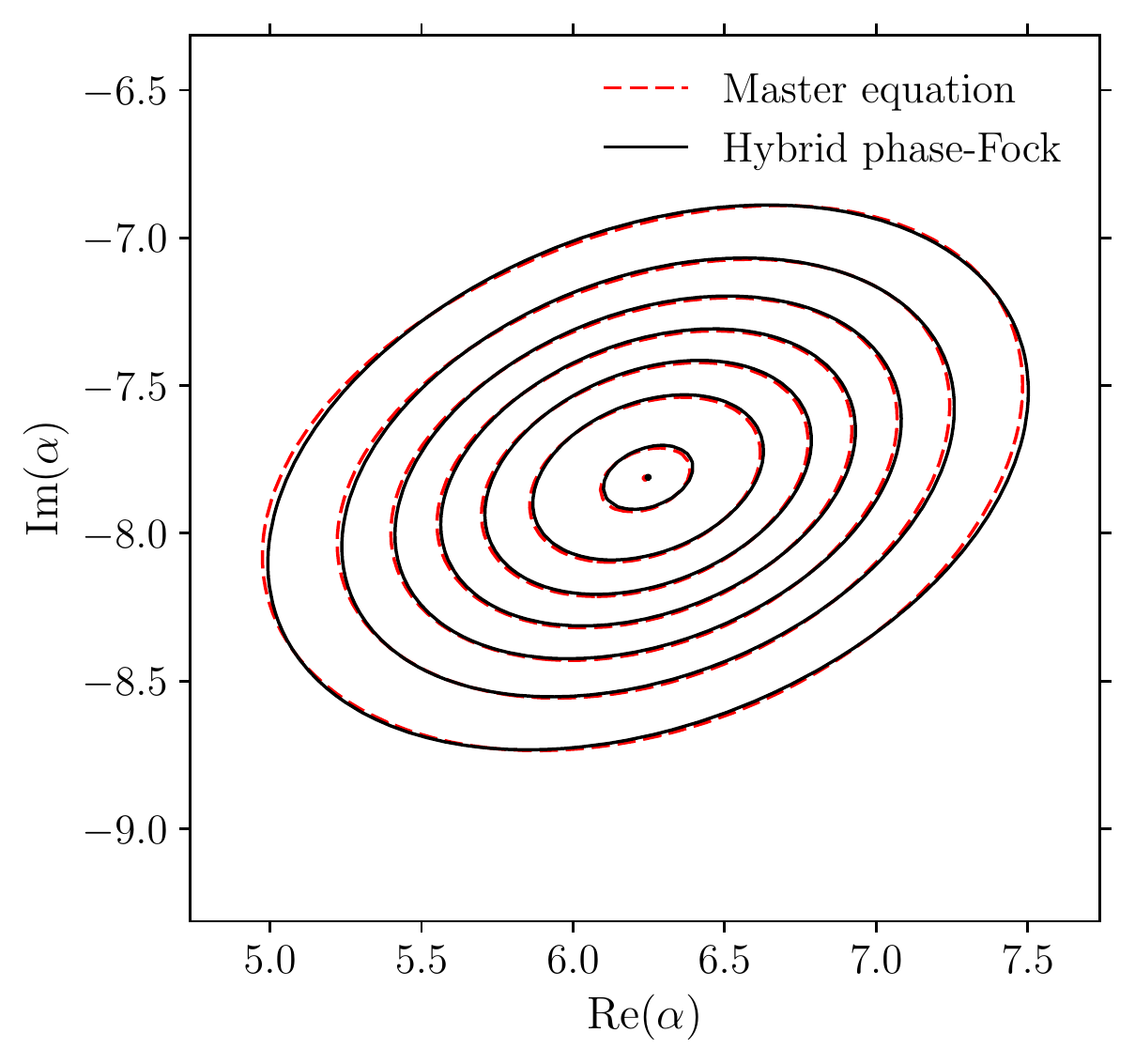}
    \end{center}
    \caption{
    Contour plot for the Wigner function $W(\alpha)$ of the resonator state.
    The black solid lines are calculated using the hybrid  evolution equations \eqref{eq:hybrid-beta}--\eqref{eq:hybrid-K}, the red dashed lines are calculated using the master equation (\ref{eq:master}).
    The parameters are the same as in Figs.\ \ref{fig:F(t)-1} and \ref{fig:bump}, the snapshot is taken at time $t=15/\kappa$.
    The contours are drawn at the levels of $1/4\pi$, $2/4\pi$, ... $7/4\pi$. The centers are indicated by black and red dots.
    }\label{fig:wigner}
\end{figure}

The main advantage of our method is a simple calculation of the resonator state deviation from a coherent state. For illustration, Fig.\ \ref{fig:wigner} shows the contour plot of the Wigner function $W(\alpha)$ of the resonator state at time moment $t=15/\kappa$ (practically the steady state) for the same parameters as in Figs.\ \ref{fig:F(t)-1} and \ref{fig:bump}. The solid black lines are calculated for our approximate hybrid-evolution state $\rho_{\rm h}$, while the dashed red lines correspond to the exact state $\rho_{\rm m}$ (at this snapshot $1-F=2.5\times 10^{-4}$).
We see that our approach gives a quite good approximation for the Wigner function; the difference is mainly because $W(\alpha)$ contour plot for the actual state $\rho_{\rm m}$ has a slightly crescent shape, while in our Gaussian-state approximation the contours are strictly elliptical. We used Eq.\  (\ref{eq:wigner-eigen}) to calculate $W(\alpha)$ for the Gaussian state $\rho_{\rm h}$, while for $\rho_{\rm m}$ we used the formula \cite{Cahill1969,Haroche2006}
    \begin{equation}\label{eq:wigner-num}
W(\alpha ) = \frac{2}{\pi}\, \text{Tr}\left[ D(-\alpha )\, \rho \, D(\alpha)\, e^{i\pi a^\dagger a}\right] ,
    \end{equation}
in which the displacement operator $D(\alpha)$ was applied numerically in the Fock space.

\begin{figure}[t]
	\begin{center}
\includegraphics[width=.95\columnwidth]{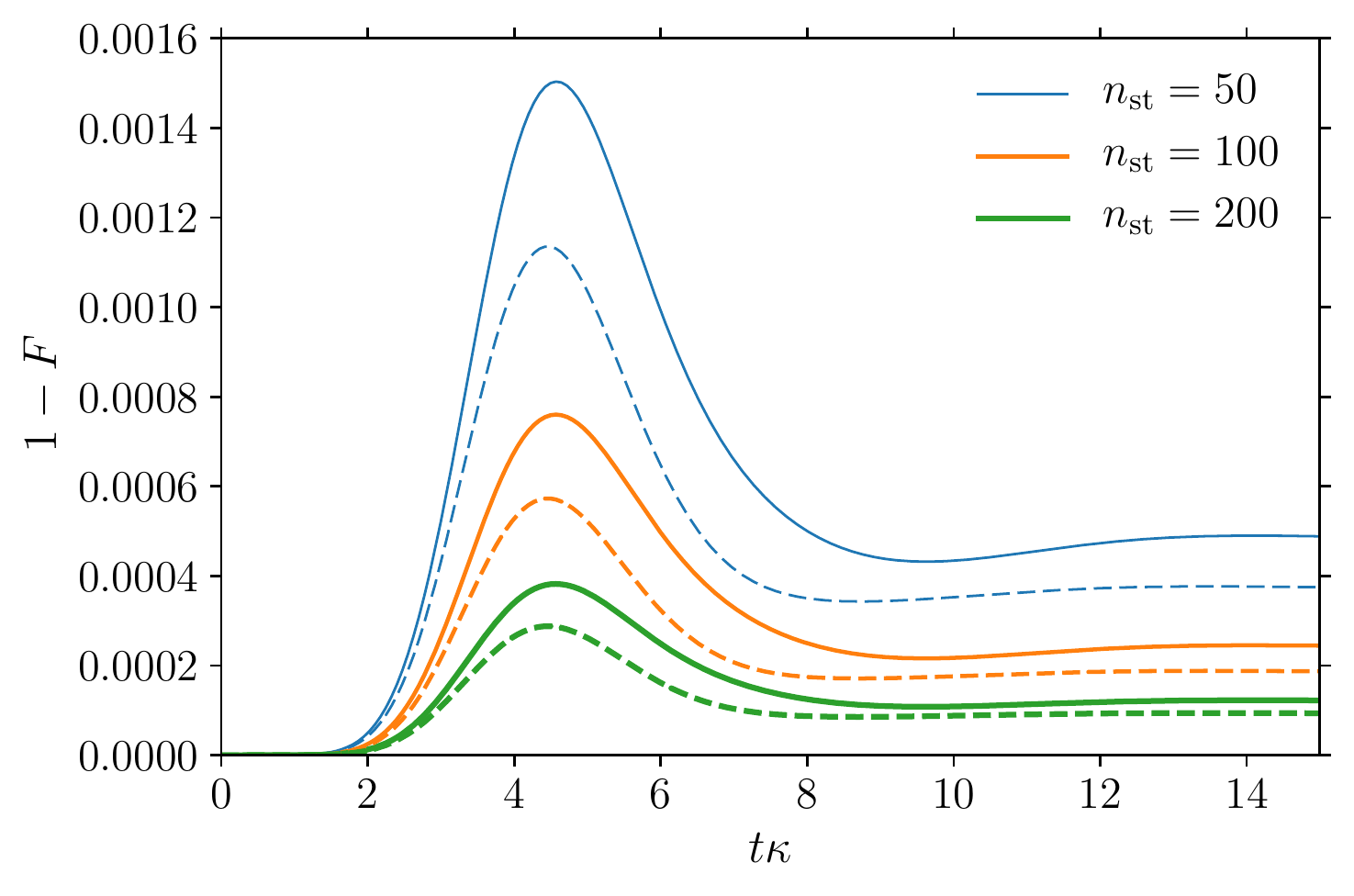}
    \end{center}
    \caption{
Solid lines: time dependence of infidelity $1-F(t)$ between the simulations based on the master equation and on our hybrid evolution equations, for the stationary-state photon numbers $n_\text{st}\approx 50$, 100, and 200 from top to bottom. The corresponding (color-matched, the same order) dashed lines show infidelity of the  Gaussian-state fit to the master-equation simulations.
The dimensionless parameters, $\varepsilon\sqrt{|\eta |}/\kappa^{3/2}=0.40$ and $(\omega_{\rm r0}-\omega_{\rm d})/\kappa =0$, are the same as in Figs.\ \ref{fig:F(t)-1}--\ref{fig:wigner}, while $\varepsilon$ and $\eta$ change from line to line (see text).
    }\label{fig:F(t)-2}
\end{figure}

Now let us check numerically the expectation that our approach should become more accurate with more photons in the resonator.
The solid lines in Fig.\ \ref{fig:F(t)-2} show the time-dependent infidelity $1-F(t)$ for the calculations using Eqs.\ \eqref{eq:hybrid-beta}--\eqref{eq:hybrid-K} (compared with the master equation results) for different number of photons.
All solid lines correspond to the same normalized drive amplitude and detuning as in Figs.\ \ref{fig:F(t)-1}--\ref{fig:wigner}: $\varepsilon\sqrt{|\eta|}/\kappa^{3/2}=0.40$ and $(\omega_{\rm r0}-\omega_{\rm d})/\kappa =0$; however, nonlinearity $\eta$ varies: from top to bottom $\eta/2\pi=-0.04$, $-0.02$, and $-0.01\,$MHz; correspondingly, the drive amplitude $\varepsilon$ also varies (with decay rate $\kappa/2\pi=5\,$MHz kept constant): $\varepsilon/2\pi=32/\sqrt{2},\, 32$, and $32\sqrt{2}$ MHz. This corresponds to the steady-state average photon number $n_{\rm st}\approx |\beta_{\rm st}|^2$ approximately equal to 50, 100, and 200 from top to bottom (note that the scaled evolution is the same as in Fig.\ \ref{fig:bump}).
As expected, the solid lines in Fig.\ \ref{fig:F(t)-2} show that the infidelity becomes smaller with more photons in the resonator. The scaling is crudely  $1-F \propto |\beta_{\rm st}|^{-2}$, as expected from Fig.\ \ref{fig:conv-beta} and Eq.\ (\ref{1-F-num}).

In addition to better accuracy, for larger $|\beta_{\rm st}|$ our approach becomes much more preferable computationally in comparison with the master-equation calculations.
As an example, for our codes (which are rather simple, Mathematica-based) the calculation of the hybrid evolution $\rho_{\rm h}(t)$ for the solid lines in Fig.\ \ref{fig:F(t)-2} took about 0.02 seconds, while obtaining the numerical master-equation solution $\rho_{\rm m}(t)$ took 0.2, 1, and 4 hours on a high-end desktop computer (longer time for larger $|\beta_{\rm st}|$). The master-equation simulation duration scales crudely quadratically with the size of the Fock space, while for our hybrid equations there is no scaling with the system size.
For the lower solid line in Fig.\ \ref{fig:F(t)-2}, our method was faster by a factor exceeding $10^5$.

Dashed lines in Fig.\ \ref{fig:F(t)-2} show infidelity of the Gaussian-state fit of $\rho_{\rm m}(t)$ for the same parameters. Comparing the solid and dashed lines, we see that most of the infidelity in our approach comes from {\it non-Gaussianity} of the actual state, thus making unimportant any possible improvements in the state center calculation by improving Eq.\ (\ref{eq:hybrid-beta}). We also see that the fraction of the infidelity coming from non-Gaussianity does not change significantly with changing number of photons.

Note that with zero initial detuning, $\omega_{\rm d}=\omega_{\rm r0}$, assumed in Figs.\ \ref{fig:F(t)-1}--\ref{fig:F(t)-2},  we automatically avoid the bistability region \cite{Landau1976,Nayfeh1995} for the steady state of a classical resonator with Kerr nonlinearity (\ref{Kerr}). Our method is generally not intended to work inside or close to this bistability region. In particular, quantum treatment formally removes the bistability \cite{Drummond1980} because of transitions due to quantum fluctuations (tunneling or quantum activation \cite{Dykman2007}), even though the rate of these transitions can be exponentially small. In contrast, our approach uses the classical equation (\ref{eq:hybrid-beta}) for the state center evolution, showing full bistability. The critical point \cite{Landau1976,Nayfeh1995} (start of the bistability) occurs at $|\tilde\varepsilon|=3^{-3/4}\approx 0.44$ and $\Delta \tilde{\omega}_{\rm d} = \sqrt{3}/2$ for the dimensionless parameters
    \begin{equation} \tilde{\varepsilon}\equiv \frac{\varepsilon \sqrt{|\eta|}}{\kappa^{3/2}}, \,\,\,\, \Delta\tilde{\omega}_{\rm d}\equiv -{\rm sign}(\eta) \, \frac{\omega_{\rm r0}-\omega_{\rm d}}{\kappa}.
    \label{tilde-epsilon}\end{equation}
For larger $|\tilde\varepsilon|$, the bistability range for $\Delta \tilde{\omega}_{\rm d}$ becomes non-zero and grows. For a given $\Delta \tilde{\omega}_{\rm d}$ above $\sqrt{3}/2$, the bistability region for the dimensionless drive amplitude is $|\tilde \varepsilon _-|\leq |\tilde \varepsilon|\leq |\tilde \varepsilon_+|$, where $|\tilde \varepsilon_\mp|^2 =\tilde{n}_\pm [\tilde{n}_\pm - \Delta \tilde{\omega}_{\rm d}]^2+ \tilde{n}_\pm /4$ and  $\tilde{n}_\pm = [2\Delta \tilde{\omega}_{\rm d} \pm \sqrt{\Delta \tilde{\omega}_{\rm d}^2-3/4}]/3 \,\,$ \cite{Drummond1980} (here $\tilde{n}$ is related to the photon number $n_{\rm st}=|\beta_{\rm st}|^2$ as $\tilde n=n_{\rm st}|\eta|/\kappa$). As mentioned above, we should avoid this bistability region when using our approach (\ref{eq:hybrid-beta})--(\ref{eq:hybrid-K}). We have checked numerically that in the vicinity of the critical point as well as near the bistability region, the unsqueezing parameter $4(D_0+b)$ may become large, indicating that our approach could become accurate only at very large number of photons.

The numerical results presented in this section show that our approach based on the hybrid evolution equations \eqref{eq:hybrid-beta}--\eqref{eq:hybrid-K} typically provides a good accuracy, which is orders of magnitude better than using the conventional approximation based on the coherent-state assumption. On the other hand, our approach is orders of magnitude faster than the full simulation based on the master equation.

\section{$3\,\text{dB}$ squeezing limit and its violation in transients}\label{sec:3dB}

Squeezing of a resonator state due to Kerr nonlinearity (\ref{Kerr}) has been discussed long ago \cite{Tanas1984,Tanas1989,Milburn1986,Kitagawa1986} (see also \cite{Khezri2016}).
A somewhat similar squeezing of the vacuum state can be produced by a parametric drive at the doubled frequency \cite{Walls2008, Drummond2004}, and in this case the steady-state squeezing of the resonator state is always less than 3 dB, i.e., $[4(D_0-b)]^{-1}\leq 2$ \cite{Milburn1981, Walls2008, Collett1984}. There were several proposals to exceed this limit in a nanomechanical system, in particular based on reservoir engineering \cite{Rabl2004, Kronwald2013}, weak measurements \cite{Ruskov2005, Szorkovszky2011}, injection of squeezed light \cite{Jahne2009}, and short optical pulses \cite{Vanner2011}. The 3 dB limit for a mechanical oscillator was recently exceeded experimentally \cite{Lei2016} by using reservoir engineering and backaction-evading measurement.

Because of a similarity \cite{Vijay2008, laflamme2011} between squeezing produced by a doubled-frequency parametric driving and by the usual non-parametric driving of a nonlinear resonator, it is natural to expect a similar 3 dB limit for squeezing in the system considered in this paper. However, we are not aware of papers, which discussed such a limit explicitly. In this section we prove that the hybrid phase-Fock-space evolution equations (\ref{eq:hybrid-beta})--(\ref{eq:hybrid-K}) indeed show the 3 dB limit for the steady-state squeezing. We also show that squeezing
may exceed this limit during the evolution.

First, let us consider squeezing in the steady state.
Substituting $\dot{W}_1=\dot{W}_2=0$ into Eqs.\ (\ref{eq:hybrid-W1}) and (\ref{eq:hybrid-W2}), we find that in the steady state
    \begin{equation}
1+16 K^2 =\frac{2W_1/\coth (\omega_{\rm r0}/2T_{\rm b}) -1}{W_1W_2}.
    \end{equation}
Therefore, from Eq.\ (\ref{eq:conv-D}) we obtain $D_0=W_1/[4W_2 \coth (\omega_{\rm r0}/2T_{\rm b})]$. Now using Eq.\ (\ref{eq:conv-b}) for the parameter $b$,  we obtain the scaled minimum quadrature variance $4(D_0-b)=W_1/[W_2 \coth (\omega_{\rm r0}/2T_{\rm b})]- \sqrt{[W_1/W_2 \coth (\omega_{\rm r0}/2T_{\rm b})]^2-W_1/W_2}$. Representing this result as
    \begin{equation}
4(D_0-b)=\frac{\coth (\omega_{\rm r0}/2T_{\rm b})}{1+\sqrt{1-\coth^2 (\omega_{\rm r0}/2T_{\rm b})\, W_2/W_1} },
    \end{equation}
we obtain  $[4(D_0-b)]^{-1}<2$ since $\coth (\omega_{\rm r0}/2T_{\rm b})\geq 1$ and $W_2/W_1$ is positive. Thus, squeezing is less than 3 dB in the steady state.

Note that the 3 dB squeezing limit can be approached only when the bath temperature $T_{\rm b}$ is zero [so that $\coth (\omega_{\rm r0}/2T_{\rm b})=1$] and when $W_1/W_2\to \infty$. Correspondingly, effective temperature $T_{\rm eff}$ becomes infinitely large because $n_{\rm th}\to \infty$, as follows from Eqs.\ (\ref{nth-W1/W2}) and (\ref{W1/W2}). We also see that in this case the maximum quadrature variance becomes infinitely large,
$4(D_0+b)\to \infty$, which indicates instability (similar to the case of reaching the 3 dB limit for parametric doubled-frequency drive \cite{Milburn1981,Walls2008}). Using Eqs.\ (\ref{eq:hybrid-beta})--(\ref{eq:hybrid-K}), we have checked numerically that 3 dB squeezing can be approached near the critical point and also near the switching point on the upper branch in the bistability region. As discussed above, our formalism is not actually intended to work in this parameter range. The hybrid equations do not have any mathematical problems in this range; however, there can be a problem with accuracy compared to the exact (master equation) evolution. In particular, when $4(D_0+b)$ becomes large near the critical point, the accuracy of the formalism requires a very large number of photons [see estimate (\ref{1-F-num})]. In addition, within the bistability region our formalism neglects switching between the quasistable states caused by fluctuations, so it can be reasonably accurate only when the switching rate is very small (that also requires a large number of photons). In spite of these issues, we can still formally use our equations, keeping in mind the potential problems.

Even simpler derivation of the 3 dB limit can be obtained using Eq.\ (\ref{eq:evol-Dpmb}). This derivation follows very closely the underlying physical idea of the derivation \cite{Walls2008, Collett1984} for the case of a parametric drive. From Eq.\ (\ref{eq:evol-Dpmb}) we find that in the steady state the  unsqueezing and inverse squeezing factors are
    \begin{equation}     4(D_0 \pm b) =\frac{\coth (\omega_{\rm r0}/2T_{\rm b})}{1 \mp 2\eta_\beta |\beta|^2 \sin (\Delta \theta)/\kappa }.
    \label{D0pmb-steady}\end{equation} Since $D_0+b>0$, there is a limitation $2\eta_\beta |\beta|^2 \sin (\Delta \theta)< \kappa$ (which is similar  to the constraint of the parametric instability). Therefore, for $4(D_0-b)$ the denominator in Eq.\  (\ref{D0pmb-steady}) is less than $2$ (and obviously positive), thus leading to the inequality $4(D_0-b)>(1/2) \coth (\omega_{\rm r0}/2T_{\rm b})\geq 1/2$.

\begin{figure}[t]
	\begin{center}
\includegraphics[width=.92\columnwidth]{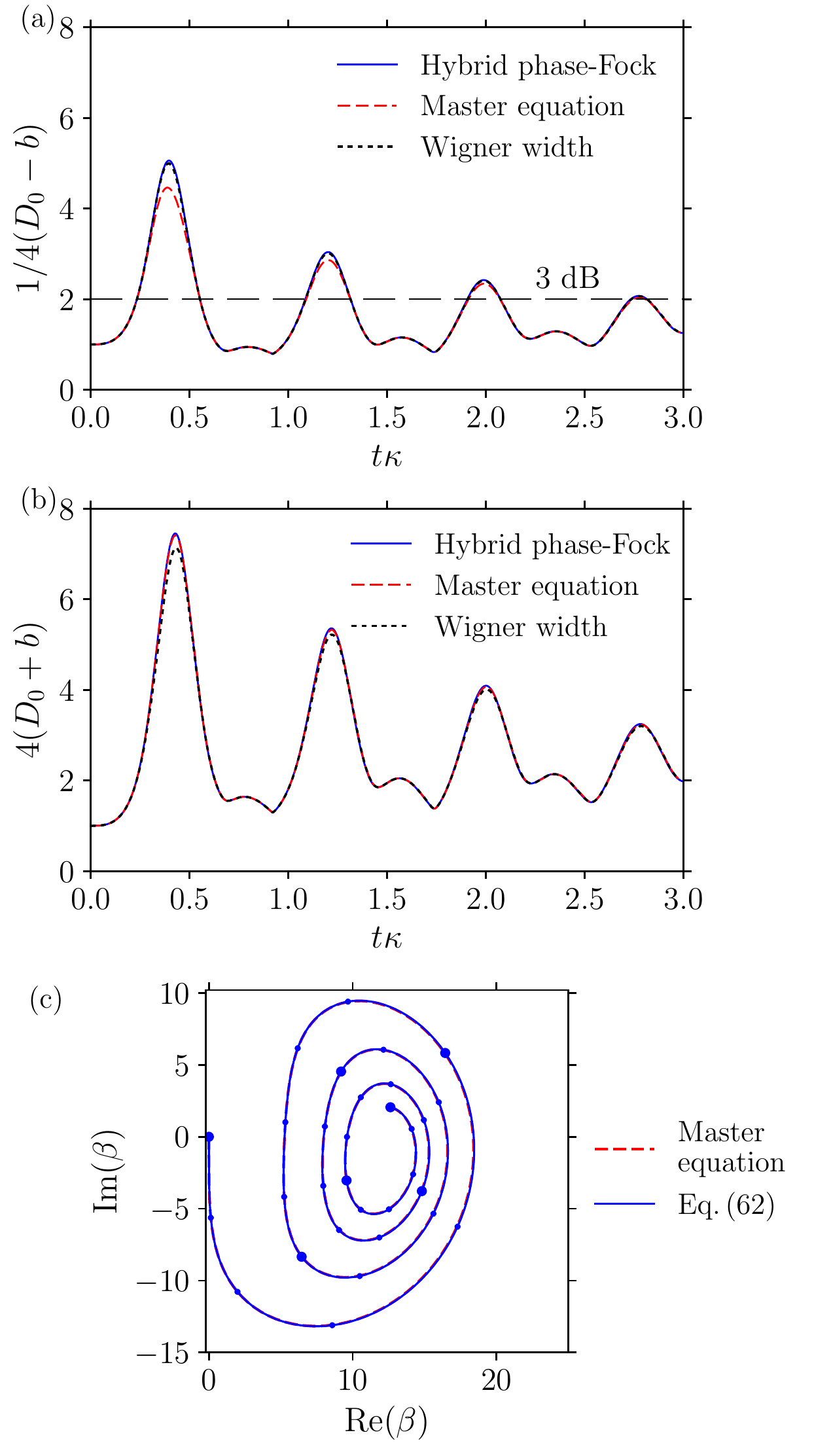}
    \end{center}
    \caption{
    Panel (a): Squeezing factor $[4(D_0-b)]^{-1}$ as a function of time $t$ for $\omega_{\rm d}/2\pi =\omega_{\rm r0}/2\pi = 6$ GHz, $\kappa/2\pi =5$ MHz, $\eta/2\pi=-0.15$ MHz, $T_{\rm b}=0$, and $\varepsilon/2\pi = 290$ MHz (so that $\varepsilon\sqrt{|\eta|}/\kappa^{3/2}=10$). The solid blue line is calculated using the hybrid evolution equations, the dashed red line is obtained from the master equation simulation, and the dotted black line is the variance of the master-equation Wigner function along the short axis. Panel (b): Unsqueezing factor $4(D_0+b)$ for the same parameters (solid blue and dashed red lines). Dotted black line is the Wigner function variance along the long axis. Panel (c): the corresponding evolution of the state center $\beta (t)$ on the phase plane. The dots are separated in time by $0.1/\kappa$, larger dots are separated by $0.5/\kappa$.
    }\label{fig:squeez}
\end{figure}

Even though the steady-state squeezing is always below 3 dB, this limit can be violated  before reaching the steady state.
As an example, the solid blue line in Fig.\ \ref{fig:squeez}(a) shows the squeezing factor $[4(D_0-b)]^{-1}$ as a function of time for the dimensionless drive amplitude $\tilde\varepsilon =\varepsilon\sqrt{|\eta|}/\kappa^{3/2}=10$, no initial detuning,  $\omega_{\rm d}=\omega_{\rm r0}$, and zero temperature of the bath. We see that  the 3 dB limit (horizontal line) is exceeded repeatedly, even though in the stationary state the squeezing is below 3 dB. This numerical result was obtained using Eqs.\ \eqref{eq:hybrid-beta}--\eqref{eq:hybrid-K}. To check it, we also performed the simulations using the master equation (\ref{eq:master}). The dashed red line in Fig.\ \ref{fig:squeez}(a) shows the corresponding result for the same parameters and $\eta/\kappa = -0.03$ (as discussed above, master equation requires more dimensionless parameters than the hybrid evolution equations); for example, this case can be realized with $\omega_{\rm d}/2\pi =\omega_{\rm r0}/2\pi = 6$ GHz, $\kappa/2\pi =5$ MHz, $\eta/2\pi=-0.15$ MHz, $T_{\rm b}=0$, and $\varepsilon/2\pi \approx 290$ MHz (these parameters can in principle be realized with a circuit QED setup by increasing the effective resonator nonlinearity $|\eta|$ using an increased qubit-resonator coupling). The maximum average number of photons in this case is approximately 350 (at $\kappa t\approx 0.4$) -- see Fig.\ \ref{fig:squeez}(c). Comparing the solid blue and dashed red lines in Fig.\ \ref{fig:squeez}(a), we see that the master equation gives a slightly smaller squeezing than the hybrid equations, but it still significantly exceeds the 3 dB value at the peaks. Note that the hybrid-equation calculation took about 0.02 seconds on a desktop computer, while the master-equation simulation took over 15 hours (the ratio of over $10^6$).

A noticeable inaccuracy of the squeezing calculation in Fig.\  \ref{fig:squeez}(a) using the hybrid equations is related to large values of the unsqueezing parameter $4(D_0+b)$ shown in Fig.\ \ref{fig:squeez}(b). At the first peak ($\kappa t\approx 0.4$) the infidelity estimate using Eq.\ (\ref{1-F-num}) for $|\beta|^2\approx 350$ gives 0.05, so we would expect a noticeable inaccuracy. We checked that the inaccuracy decreases with decreasing nonlinearity $|\eta|/\kappa$ while keeping $\varepsilon \sqrt{|\eta|}/\kappa^{3/2}$ fixed; this increases the number of photons, which scales as $\kappa/|\eta|$. (Since further increase of the photon number is very difficult for the master-equation simulations, we actually checked that the inaccuracy in Fig.\ \ref{fig:squeez}(a) increases with decreasing number of photons by  increasing $|\eta|/\kappa$.) Note that the unsqueezing parameters calculated by the hybrid equations and by the master equation [solid blue and dashed red lines in Fig.\ \ref{fig:squeez}(b)] practically coincide with each other.

Figure  \ref{fig:squeez}(c) shows the evolution of the state center $\beta(t)$ on the phase plane, with dots separated in time by  $0.1/\kappa$ (larger dots are separated by $0.5/\kappa$); the results from Eq.\ (\ref{eq:hybrid-beta}) and master equation practically coincide with each other. Comparing Fig.\  \ref{fig:squeez}(c)  with Figs.\ \ref{fig:squeez}(a) and \ref{fig:squeez}(b), we see that peaks in squeezing and unsqueezing approximately correspond to maxima of the photon number $|\beta |^2$. The minima of the photon number correspond to small bumps on the lines in Figs.\ \ref{fig:squeez}(a) and \ref{fig:squeez}(b).

We expect that the difference between the solid blue and dashed red lines for the squeezing factor in Fig.\ \ref{fig:squeez}(a) can be mostly explained by a non-Gaussian shape of the actual states produced by the master equation. This non-Gaussianity can be seen as a slightly crescent shape of the Wigner function in the phase plane (see Fig.\ \ref{fig:wigner}), with slightly curved ``arms'' along the long axis, instead of the perfect elliptical shape. However, the  bending of the ``arms'' produces a smaller effect along the short axis.
To check this hypothesis, we have calculated the Wigner function variance along the short axis by numerically fitting the master-equation  Wigner function along the short axis (passing through the state center) with a one-dimensional Gaussian model. The result is shown by the dotted black line in Fig.\ \ref{fig:squeez}(a). It is almost indistinguishable from the blue solid line, thus confirming that squeezing calculated by our hybrid-evolution method is essentially the squeezing of the Wigner function along the short axis (which is slightly different from the usual ``integrated'' definition based on the quadrature variance, which is affected by bending of the ``arms''). In contrast, the Wigner function variance along the long axis, shown by black dotted line in Fig.\ \ref{fig:squeez}(b), noticeably differs from the quadrature variance shown by the solid blue (or dashed red) line. This is expected because the Wigner function along the long axis is significantly more affected by bending of the ``arms''.

\begin{figure}[t]
	\begin{center}
\includegraphics[width=.92\columnwidth]{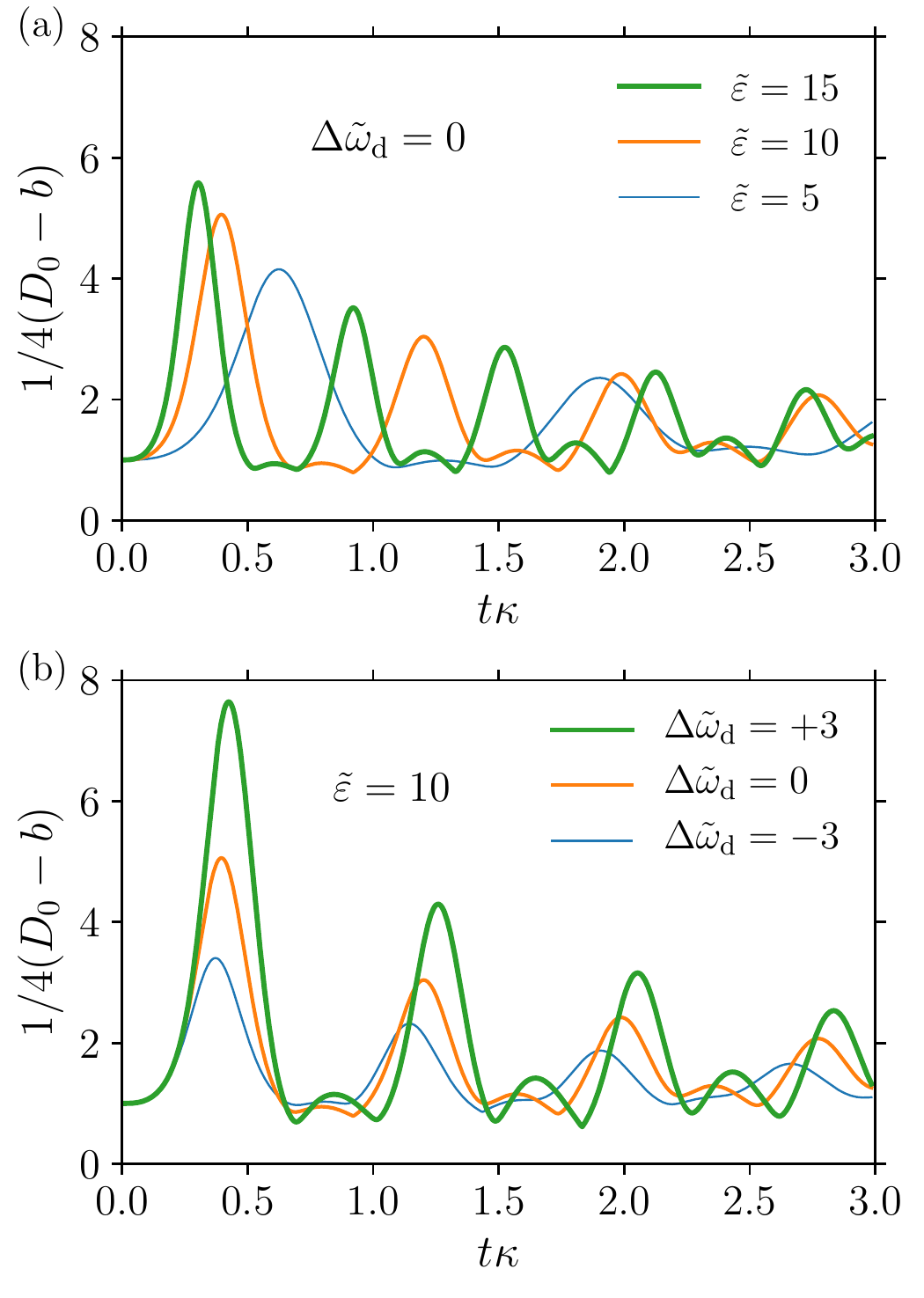}
    \end{center}
    \caption{
Time dependence of the squeezing factor $[4(D_0-b)]^{-1}$, calculated using the hybrid evolution equations  (\ref{eq:hybrid-beta})--(\ref{eq:hybrid-K}). The lines in panel (a) are for zero initial detuning, $\omega_{\rm d}=\omega_{\rm r0}$, zero bath temperature, $T_{\rm b}=0$, and dimensionless drive amplitudes $\varepsilon\sqrt{|\eta|}/\kappa^{3/2}=15$, $10$, and $5$ (from top to bottom). The lines in panel (b) are for $\varepsilon\sqrt{|\eta|}/\kappa^{3/2}=10$, $T_{\rm b}=0$, and dimensionless initial detunings $(\omega_{\rm d}-\omega_{\rm r0})/\kappa \,{\rm sign}(\eta)=3$, $0$, and $-3$ (from top to bottom). All lines repeatedly exceed the 3 dB squeezing limit (factor of 2).
    }\label{fig:over3dB}
\end{figure}

Figure \ref{fig:over3dB} shows time-dependence of the squeezing factor $[4(D_0-b)]^{-1}$ for various parameters; these results are obtained using the hybrid equations  \eqref{eq:hybrid-beta}--\eqref{eq:hybrid-K}. In Fig.\ \ref{fig:over3dB}(a) we assume zero initial detuning and zero bath temperature, $\omega_{\rm d}=\omega_{\rm r0}$, $T_{\rm b}=0$, while varying the dimensionless drive amplitude, $\tilde{\varepsilon}\equiv\varepsilon\sqrt{|\eta|}/\kappa^{3/2}=5$, $10$, and $15$. In Fig.\  \ref{fig:over3dB}(b) we keep the amplitude fixed,  $\tilde\varepsilon =10$, and vary the detuning, $\Delta \tilde{\omega}_{\rm d}\equiv {\rm sign}(\eta)(\omega_{\rm d}-\omega_{\rm r0})/\kappa=-3$, $0$, and $3$ (the temperature is still zero). We see that a larger  squeezing can be achieved with a larger amplitude of the drive and also with a detuning, which moves the operating point closer to the bistability region (for $\tilde{\varepsilon}=10$ the bistability region starts at $\Delta \tilde\omega_{\rm d}=8.75$). Note that a larger squeezing also leads to a larger unsqueezing $4(D_0+b)$; for example, the maximum squeezing factor of 5.6 in Fig.\ \ref{fig:over3dB}(a) for $\tilde{\varepsilon}=15$ corresponds to $4(D_0+b)=7.8$ (at this point $|\beta|^2=13.9\, \kappa/\eta$). Similarly, the maximum squeezing factor of 7.6 in Fig.\ \ref{fig:over3dB}(b) for $\Delta \tilde{\omega}_{\rm d}=3$ corresponds to $4(D_0+b)=16.3$ (at this point $|\beta|^2=14.1\, \kappa/\eta$). This means that to observe these large values of squeezing, we would need very many photons in the resonator. From Eq.\ (\ref{1-F-num}) and numerical results in Sec.\ \ref{subsec:fid-hybrid}, we expect that validity of our formalism requires
    \be
    \bar{n} \approx |\beta|^2  \gg [4(D_0+b)]^3 .
    \label{eq:validity}\ee
Therefore, we estimate that for the upper (green) lines in Figs.\ \ref{fig:over3dB}(a) and \ref{fig:over3dB}(b) to be reasonably accurate, we need over 500 and 4,000 photons, respectively. Therefore, we cannot check results of Fig.\ \ref{fig:over3dB} against the master equation. However, since the results of the hybrid equations and the master equation agree well with each other in the range where the master equation requires reasonable computational resources, we believe that our Eqs.\ (\ref{eq:hybrid-beta})--(\ref{eq:hybrid-K}) can still be reliably used for parameters when the master equation already cannot be used because of too large Hilbert space.

\section{Conclusion}\label{sec:conclusion}

In this paper we have introduced a new approximate method for numerical calculation of quantum evolution of a weakly nonlinear resonator due to drive and dissipation. This method is most accurate for large number of photons in the resonator (hundreds, thousands or more). This is exactly the regime where the conventional method based on the master equation becomes inapplicable because of too large Hilbert space. For a few hundred photons in the resonator (when the master equation can still be used), our method is faster by a factor of over $10^5$, while providing a very good accuracy.

The method is based on a hybrid description of a quantum state, which uses both phase-space and Fock-space parameters. The advantage is that evolution due to drive and dissipation can be naturally described in the phase space, while evolution due to nonlinearity has a simple description in the Fock space. We combined both descriptions by proving that a phase-space Gaussian state with many photons has a simple approximate representation in the Fock space, Eq.\ (\ref{eq:FG}), which is also Gaussian. Thus, our method essentially uses the Gaussian-state approximation for an evolving quantum state. It is not applicable for quantum dynamics involving cat-states, but is well-applicable for analyzing squeezing, unsqueezing, and effective heating of the resonator state due to weak nonlinearity.

The method describes the quantum evolution via solving four ordinary differential equations, Eqs.\ (\ref{eq:hybrid-beta})--(\ref{eq:hybrid-K}). One of them, Eq.\ (\ref{eq:hybrid-beta}), is decoupled from other equations and describes the  evolution of the state center $\beta(t)$ on the (complex) phase plane. This is the usual classical equation, which takes into account resonator nonlinearity. (This equation can be generalized by coupling it with other equations; however, in our numerical analysis we did not find a significant improvement of accuracy by doing this.) Other three equations, Eqs.\ (\ref{eq:hybrid-W1})--(\ref{eq:hybrid-K}),  essentially describe evolution of the three quantum parameters of a Gaussian state (maximum and minimum quadrature variances $D_0\pm b$ and the short-axis angle $\theta/2$ on the phase plane); however, this is done using the Fock-space parameters ($W_1$, $W_2$, and $K$). For conversion of the results into the phase-space description we use Eqs.\ (\ref{eq:conv-D})--(\ref{eq:conv-theta}). It is also possible to use Eqs. (\ref{eq:hybrid-D0})--(\ref{eq:hybrid-theta}) to simulate evolution of the parameters $D_0$, $b$, and $\theta$ directly, though in this paper we have not focused on this way of analysis. Physically, our approach is related to linearization of fluctuations around a classical trajectory \cite{Ludwig1975}; however, formally it is based on a different framework.

Numerical accuracy of our method has been studied in Sec.\ \ref{sec:numerical}. Somewhat surprisingly, it works well not only for a very large number of photons (as expected), but may also provide a reasonable accuracy when there are only a few dozen photons in the resonator. It is important that the method accurately describes the evolution starting with vacuum (where it formally should not work); this is because during the evolution, effects of nonlinearity become important at larger number of photons where the method already works well.

The method becomes inaccurate when a quantum state cannot be reasonably represented as a Gaussian state. In our simulations this has been usually the case when the long-axis quadrature variance $D_0+b$ is large, while the number of photons $|\beta|^2$ is not sufficiently large, so that the Wigner function of the state has a noticeable crescent shape in the phase plane. We have found numerically that Eq.\ (\ref{eq:validity}) can be used for a crude estimate of the applicability range of the method; a weaker condition, $|\beta|^2 > [4(D_0+b)]^3$, still provides a reasonably good accuracy. Because of a growing inaccuracy, the method is not intended to be used close to the critical point of the resonator bistability, where the long-axis quadrature variance $D_0+b$ becomes large. Similarly, the method is not intended to be used within the bistability region, since it neglects switchings between the quasistable states caused by fluctuations. Nevertheless, the equations of the method can be formally used in any regime, keeping in mind these reasons for potential inaccuracy of the results compared with full master-equation simulations. We have checked (Appendix \ref{app:steady}) that our analytical results for the steady state agree with the results of Refs.\ \cite{Drummond1980} and \cite{Dykman-2012}.

As an example, In Sec.\ \ref{sec:3dB} the equations of our method have been used to derive the 3 dB limit for the steady-state squeezing of a pumped and damped weakly nonlinear resonator. We have also shown numerically that squeezing during transients can significantly exceed this 3 dB limit (Fig.\ \ref{fig:over3dB}).
We emphasize that such an analysis is very difficult using the master equation because a large squeezing typically requires large number of photons in the resonator and therefore large Hilbert space. In contrast, our calculations take only a fraction of a second, independently of the photon number.

We hope that our method can be useful in various fields of research involving squeezing of weakly nonlinear resonators with large number of quantum excitations. In particular, it can be useful for circuit QED systems, in which a weak resonator nonlinearity is induced by interaction with a qubit. Note that our method describes squeezing of the resonator state, but it is not directly applicable to a transmitted/reflected microwave field outside of the resonator (such generalization can be a subject of future research). Our method can also be useful in analysis of nanomechanical systems at low temperatures.

\begin{acknowledgements}
We would like to thank Juan Atalaya, Aashish Clerk, Justin Dressel,  and Mark Dykman for useful discussions.
The work was supported by ARO grant W911NF-15-1-0496.
\end{acknowledgements}

\appendix

\section{Rotating-frame evolution of a~linear-resonator state}\label{app:rf}

In this appendix, we discuss derivation of the rotating-frame equations \eqref{eq:evol-beta}--\eqref{eq:evol-theta} for evolution of the Gaussian-state parameters $\beta$, $D_0$, $b$, and $\theta$ from the laboratory-frame equations (\ref{eq:evol-x})--(\ref{eq:evol-Dxp}), using the rotating wave approximation (RWA).

Let us start with introducing the rotating frame based on the drive frequency $\omega_{\rm d}$,  by defining the dimensionless rotating-frame position and momentum operators $\hat{\tilde{x}}$ and $\hat{\tilde{p}}$ as
    \begin{equation}\label{eq:tilde(a)}
	\hat{\tilde{x}}+i\hat{\tilde{p}} = (\hat{x} + i\hat{p})\, e^{i\omega_{\rm d}t}.
    \end{equation}
This is equivalent to introducing a new lowering operator $\hat{\tilde{a}}=\hat{a}\, e^{i\omega_{\rm d}t}$.
From Eq.\,\eqref{eq:tilde(a)} we obtain the canonical transformation
    \begin{align}
	\hat{x} &= \hat{\tilde{x}}\cos\omega_{\rm d}t + \hat{\tilde{p}}\sin\omega_{\rm d}t, \label{eq:tilde(x)} \\
	\hat{p} &= \hat{\tilde{p}}\cos\omega_{\rm d}t - \hat{\tilde{x}}\sin\omega_{\rm d}t. \label{eq:tilde(p)}
    \end{align}

To find the rotating-frame evolution equation for the Gaussian state center, we use Eqs.\ \eqref{eq:evol-x} and \eqref{eq:evol-p} for the evolution of $x_{\rm c}=\langle \hat{x}\rangle$ and $p_{\rm c}=\langle \hat{p}\rangle$, and convert them into equations for $\tilde{x}_{\rm c}=\langle \hat{\tilde{x}}\rangle$ and $\tilde{p}_{\rm c}=\langle \hat{\tilde{p}}\rangle$, thus obtaining
    \begin{align}
& \frac{d}{dt}(\tilde{x}_{\rm c} + i\tilde{p}_{\rm c}) = -i(\omega_{\rm r}-\omega_{\rm d})(\tilde{x}_{\rm c}+i\tilde{p}_{\rm c})
  -i\varepsilon -i\varepsilon^*e^{i2\omega_{\rm d}t}
    \nonumber \\
& \hspace{1.5cm} -i\kappa\left( \tilde{p}_{\rm c}\frac{1+e^{i2\omega_{\rm d}t}}{2} + \tilde{x}_{\rm c}\frac{1-e^{i2\omega_{\rm d}t}}{2i} \right) .
    \end{align}
This equation is still exact.
Now using RWA, we neglect the terms oscillating with frequency $2\omega_{\rm d}$, thus obtaining slow evolution of the Gaussian state center,
    \begin{equation}
	\dot{\beta} = -i(\omega_{\rm r}-\omega_{\rm d})\beta - \frac{\kappa}{2}\beta -i\varepsilon, \,\,\, \beta \equiv \tilde{x}_{\rm c}+i\tilde{p}_{\rm c} ,
    \end{equation}
which is Eq.\ (\ref{eq:evol-beta}).

To derive Eqs.\ (\ref{eq:evol-D0}) and (\ref{eq:evol-b}) for $\dot{D}_0$ and $\dot{b}$, let us start with expressing $D_x$, $D_p$, and $D_{xp}$ via the corresponding rotating-frame quantities $D_{\tilde{x}}$, $D_{\tilde{p}}$, and $D_{\tilde{x}\tilde{p}}$ (with obvious definitions)
    \begin{eqnarray}
&& \hspace{-0.0cm} D_x = D_{\tilde{x}}\cos^2(\omega_{\rm d}t) + D_{\tilde{p}}\sin^2(\omega_{\rm d}t) + D_{\tilde{x}\tilde{p}}\sin (2\omega_{\rm d}t), \qquad
    \label{eq:tilde(Dx)}\\
&& \hspace{-0.0cm} D_p = D_{\tilde{x}}\sin^2(\omega_{\rm d}t) + D_{\tilde{p}}\cos^2(\omega_{\rm d}t) - D_{\tilde{x}\tilde{p}}\sin (2\omega_{\rm d}t),
    \label{eq:tilde(Dp)} \\
&& \hspace{-0.0cm} D_{xp} = D_{\tilde{x}\tilde{p}}\cos (2\omega_{\rm d}t) + (1/2) (D_{\tilde{p}} - D_{\tilde{x}}) \sin (2\omega_{\rm d}t) .
    \label{eq:tilde(Dxp)}
    \end{eqnarray}
Note that $D_0 \equiv (D_x + D_p)/2$ has the same expression in the rotating frame, $D_0 =  (D_{\tilde{x}} + D_{\tilde{p}})/2$; similarly,  $b^2 \equiv  (D_p-D_x)^2/4 + D_{xp}^2$ can also be expressed as $b^2=  (D_{\tilde{p}}-D_{\tilde{x}})^2/4 + D_{\tilde{x}\tilde{p}}^2$.

For the evolution of $D_0$, from Eqs.\,\eqref{eq:evol-Dx} and \eqref{eq:evol-Dp} we find 	$ \dot{D}_0 = -\kappa D_p + (\kappa /4) \coth (\omega_{\rm r}/2T_{\rm b})$.
Then using Eq.\ \eqref{eq:tilde(Dp)}, we obtain
    \begin{align}
	\dot{D}_0 =& -\kappa [ D_{\tilde{x}}\sin^2(\omega_{\rm d}t) + D_{\tilde{p}}\cos^2(\omega_{\rm d}t) - D_{\tilde{x}\tilde{p}}\sin (2\omega_{\rm d}t )]
    \nonumber \\
	& +(\kappa /4) \coth (\omega_{\rm r}/2T_{\rm b}) .
    \end{align}
Now using RWA, we neglect the terms oscillating with frequency $2\omega_{\rm d}$, so that $\sin^2(\omega_{\rm d}t) \to 1/2$, $\cos^2(\omega_{\rm d}t) \to 1/2$, and $\sin (2\omega_{\rm d}t)\to 0$.
This gives us
    \begin{equation}
	\dot{D_0} = -\kappa D_0 + (\kappa /4) \coth (\omega_{\rm r}/2T_{\rm b}) ,
    \end{equation}
which is Eq.\ (\ref{eq:evol-D0}).

For the evolution of $b$, from  Eqs.\,\eqref{eq:evol-Dx}--\eqref{eq:evol-Dxp} we obtain
    \begin{align}
	 d (b^2)/dt = & \, (\kappa /4) (D_p-D_x) \coth (\omega_{\rm r}/2T_{\rm b})
\nonumber \\
& - \kappa (D_p-D_x) D_p  - 2\kappa D_{xp}^2.
    \label{eq:dot(b2)}
    \end{align}
Within RWA, the first term on the right-hand side is zero because $D_p-D_x$ oscillates with frequency $2\omega_{\rm d}$ [see Eqs.\ \eqref{eq:tilde(Dx)} and \eqref{eq:tilde(Dp)}].
The second term is not zero because $D_p$ has also a part oscillating with $2\omega_{\rm d}$; averaging over these oscillations we obtain $-\kappa [ D_{\tilde{x}\tilde{p}}^2 + (D_{\tilde{p}}-D_{\tilde{x}})^2 /4]$, which equals $-\kappa b^2$.
Similarly, for the third term we use Eq.\ (\ref{eq:tilde(Dxp)}) and averaging over the oscillations obtain $-\kappa [ D_{\tilde{x}\tilde{p}}^2 + (D_{\tilde{p}}-D_{\tilde{x}})^2 /4]$, which is again $-\kappa b^2$.
Thus, within RWA
    \begin{equation}
	 d (b^2)/dt = -2\kappa b^2.
    \end{equation}
Equivalently, $\dot{b} = -\kappa b$, which is Eq.\ (\ref{eq:evol-b}).

To derive Eq.\ (\ref{eq:evol-theta}) for $\dot{\theta}$, we start with Eqs.\ (\ref{eq:Theta-1}) and (\ref{Theta-theta}), which give
    \begin{align}
& \theta = \arctan\left(\frac{2D_{xp}}{D_x-D_p}\right) + 2\omega_{\rm d} t
    \nonumber \\
& \hspace{0.4cm} + (\pi/2) [1+{\rm sign} (D_x-D_p)] .
    \end{align}
Neglecting the last term, the time derivative is
    \begin{equation}
	\dot{\theta} = \frac{\dot{D}_{xp}(D_x-D_p)-D_{xp}(\dot{D}_x-\dot{D}_p)}{2b^2} + 2\omega_{\rm d}.
    \label{theta-dot-app}
    \end{equation}
Using Eqs.\,\eqref{eq:evol-Dx}--\eqref{eq:evol-Dxp}, we find that the numerator here is $-4\omega_{\rm r}b^2-2\kappa D_{xp}D_0+(\kappa /2) D_{xp}\coth(\omega_{\rm r}/2T_{\rm b})$, in which the only non-oscillating term is $-4\omega_{\rm r}b^2$.
Dividing it by $2b^2$ and adding $2\omega_{\rm d}$, from Eq.\ (\ref{theta-dot-app}) we obtain $\dot{\theta}=-2(\omega_{\rm r}-\omega_{\rm d})$, which is Eq.\ (\ref{eq:evol-theta}).

\section{Equivalence between Gaussian and Fock-space Gaussian states}\label{app:conversion}

In this appendix, we show that the Fock-space Gaussian state introduced in Eq.\,\eqref{eq:FG} is approximately the same as the standard Gaussian state [Eq.\ (\ref{eq:W-norm})] in the limit of large photon number, $|\beta | \gg 1$, and derive the conversion relations (\ref{eq:conv-beta})--(\ref{eq:conv-theta}).
This is done by comparing the Husimi $Q$-functions of the Gaussian and Fock-space Gaussian states.
We use the rotating frame and characterize the Gaussian state by the complex parameter $\beta$ (center) and three real parameters: $D_0$, $b$, and $\theta$ -- see Eqs.\ (\ref{eq:D-b})--(\ref{eq:Theta-1}).
The Fock-space Gaussian state is characterized by the complex parameter $e^{i\phi_\beta}|\beta|$ (which is chosen to be the same as $\beta$) and three real parameters: $W_1$, $W_2$, and $K$ -- see Eq.\ (\ref{eq:FG}).

The Husimi $Q$-function $Q(\alpha )$ of a state with density matrix $\rho$ is defined via  its overlap with the coherent state $|\alpha\rangle$,
    \begin{equation}\label{Q-def-app}
	Q(\alpha )=\frac{1}{\pi}\, \langle \alpha | \rho |\alpha\rangle , \,\,\, |\alpha \rangle =e^{-\frac{1}{2}\,|\alpha|^2} \sum_{n=0}^\infty \frac{\alpha^n}{\sqrt{n!}}\, |n\rangle ,
    \end{equation}
where $\alpha =\tilde{x}+i\tilde{p}$ assumes the rotating frame, in contrast to the notation  $\alpha$ used in Sec.\ \ref{subsec:Gaussian-review}.
The function $Q(\alpha)$ can be calculated from the Wigner function $W(\alpha)$ (here in the rotating frame; note a slightly different notation used in Sec.\ \ref{subsec:Gaussian-review}),
    \begin{equation}
	Q(\alpha)=\frac{2}{\pi} \int W (\alpha ') \, e^{-2|\alpha -\alpha'|^2}\, d{\rm Re}(\alpha')\, d{\rm Im}(\alpha').
    \end{equation}
For the Gaussian state (\ref{eq:W-norm}) it is equal
    \begin{eqnarray}\label{eq:Q-gauss}
&& Q(\alpha ) = \pi^{-1} \left[4(D_0-b+1/4)(D_0+b+1/4)\right]^{-1/2}
    \nonumber \\
&& \hspace{0.5cm} \times \exp \bigg\{ -\frac{(D_0+b\cos\theta +1/4) \, [{\rm Re} (\alpha -\beta )]^2}{2(D_0-b+1/4)(D_0+b+1/4)}
    \nonumber \\
&& \hspace{1.7cm}  -\frac{(D_0-b\cos\theta +1/4) \, [\text{Im} (\alpha -\beta)]^2 }{2(D_0-b+1/4)(D_0+b+1/4)}\,
    \nonumber \\
&& \hspace{1.7cm} -\frac{(2b\sin\theta) \, {\rm Re} (\alpha -\beta ) \, {\rm Im} (\alpha -\beta )
}{2(D_0-b+1/4)(D_0+b+1/4)}  \bigg\} .
    \end{eqnarray}
Recall that $\beta$ is the Gaussian state center, $D_0+b$ is the maximum quadrature variance, $D_0-b$ is the minimum quadrature variance, and $\theta /2$ is the angle between the minimum quadrature direction and $\tilde{x}$-axis (see Fig.\ \ref{fig:quad}).
Note that in the diagonal basis, Eq.\ (\ref{eq:Q-gauss}) reduces to Eq.\ (\ref{eq:Q-gaussian-eigen}), up to a slight change of notations.

Now let us calculate the $Q$-function for the Fock-space Gaussian state, Eq.\ (\ref{eq:FG}), and compare it with Eq.\ (\ref{eq:Q-gauss}).
We will use a series of approximations to calculate $Q(\alpha)$.
First, for $|\beta| \gg 1$ we can also assume $|\alpha |\gg 1$; then the coherent state $|\alpha\rangle$ in Eq.\ (\ref{Q-def-app}) can be approximated as $|\alpha\rangle \approx (2\pi|\alpha|^2)^{-1/4}\sum_n \exp[-(n-|\alpha|^2)^2/4|\alpha|^2]\exp[i n \phi_\alpha ]$, where $\phi_\alpha ={\rm arg}(\alpha )$, so that the $Q$-function is approximately
\begin{eqnarray}
&& Q(\alpha) = \frac{1}{\pi \sqrt{2\pi|\alpha|^2} } \sum_{n,m=0}^\infty \rho_{nm}
\exp \bigg[  -\frac{(n-|\alpha|^2)^2}{4|\alpha|^2}
    \nonumber\\
&& \hspace{2.5cm} -\frac{(m-|\alpha|^2)^2}{4|\alpha|^2} -i\phi_{\alpha}(n-m) \bigg] .
    \label{eq:Q}
\end{eqnarray}
Substituting $\rho_{nm}$ from Eq.\ (\ref{eq:FG}), we obtain
\begin{eqnarray}
&& Q(\alpha) = N \sum\nolimits_{n,m} \exp[-A n^2- \tilde{A} m^2  - B(m)\, n
    \nonumber \\
&& \hspace{3.4cm}
- \tilde{B} m -C],
    \label{Q-FG-1-app}\\
&& N= \pi^{-1}(4\pi^2W_1|\beta|^2|\alpha|^2)^{-1/2},
    \label{N-def} \\
&& A= \frac{1}{4|\alpha|^2} +\frac{1}{8W_1|\beta|^2}+\frac{1}{8W_2|\beta|^2}+i\frac{K}{|\beta|^2},
      \label{A-def}\\
&& \tilde{A} = \frac{1}{4|\alpha|^2}+\frac{1}{8W_1|\beta|^2}+\frac{1}{8W_2|\beta|^2}-i\frac{K}{|\beta|^2},
      \label{tilde-A-def}\\
&& B(m)= -\frac{1}{2}+i(\phi_{\alpha}-\phi_\beta) +\frac{m}{4W_1|\beta|^2}-\frac{1}{2W_1}
    \nonumber \\
&& \hspace{1.3cm} -\frac{m}{4W_2|\beta|^2}-2iK ,
      \label{B(m)-def}\\
&& \tilde{B}=  -\frac{1}{2}-i(\phi_{\alpha}-\phi_\beta)-\frac{1}{2W_1}+2iK ,
      \label{tilde-B-def}\\
&& C= \frac{|\alpha|^2}{2}+\frac{|\beta|^2}{2W_1}.
      \label{C-def}
\end{eqnarray}
Then replacing summation over $n$ and $m$ by integration within infinite limits (assuming $|\beta|\gg 1$) and calculating the integral over $n$, we find
\begin{align}
	Q (\alpha) = N \frac{\sqrt{\pi}e^{-C}}{\sqrt{A}}\int\limits_{-\infty}^\infty  \exp \bigg[ \frac{[B(m)]^2}{4A} -\tilde{A}m^2-\tilde{B}m \bigg] dm .
    \label{eq:q-mid}
\end{align}
Using Eq.\ (\ref{B(m)-def}), we then represent $[B(m)]^2/4A$  as
\begin{eqnarray}
&& [B(m)]^2/4A = \bar{A} m^2 + \bar{B} m + \bar{C},
    \\
&& \bar{A} = \frac{1}{4A} \Big( \frac{1}{4W_1|\beta|^2} \Big) ^2 \Big( 1-\frac{W_1}{W_2}\Big) ^2 ,
    \\
&& \bar{B} = \frac{1}{4A}\, \frac{1}{2W_1|\beta|^2} \Big( 1-\frac{W_1}{W_2}\Big) \Big[ -\frac{1}{2}-\frac{1}{2W_1}
    \nonumber \\
&& \hspace{3.5cm} +i(\phi_{\alpha}-\phi_\beta)-2iK\Big] , \qquad
    \\
&&  \bar{C} = \frac{1}{4A} \Big( -\frac{1}{2} -\frac{1}{2W_1} +i(\phi_{\alpha}-\phi_\beta) -2iK\Big)^2 .
\end{eqnarray}
Then the exponent in Eq.\ (\ref{eq:q-mid}) is $\exp[-(\tilde{A}-\bar{A})m^2 - (\tilde{B}-\bar{B})m]$ and its integral over $dm$ can be easily calculated,
\begin{eqnarray}
&& Q(\alpha) = (2\pi\sqrt{W_1}|\beta||\alpha|)^{-1} [A(\tilde{A}-\bar{A})]^{-1/2}
     \nonumber \\
&& \hspace{1.2cm} \times \exp \{ (\tilde{B}-\bar{B})^2/[4(\tilde{A}-\bar{A})] -C-\bar{C} \} .  \qquad
    \label{eq:Q-full}
\end{eqnarray}

Since we want to compare this result with Eq.\ (\ref{eq:Q-gauss}), we need to find its dependence on the difference $\alpha-\beta$.
Assuming $|\beta|\gg 1$, we expand Eq.\ (\ref{eq:Q-full}) up to second order in ${\rm Re}(\alpha-\beta )$ and  ${\rm Im}(\alpha-\beta )$.
Let us consider first the special case when $\beta$ is real ($\beta >0$), so that $\phi_\beta =0$.
Then expansion of Eq.\ (\ref{eq:Q-full}) produces (after some algebra) the result
\begin{eqnarray}
&& Q (\alpha) \approx  \frac{1}{\pi} \Big( \frac{1}{4} +\frac{W_1}{4W_2} +\frac{1}{4W_2} +\frac{W_1}{4} +4K^2W_1 \Big)^{-1/2}
    \nonumber \\
&& \hspace{0.7cm} \times \exp \bigg\{ -\frac{2(1+W_2+16W_1W_2K^2)\, [{\rm Re}(\alpha-\beta)]^2} {1+W_1+W_2+W_1W_2(1+16K^2)}
    \nonumber \\
&& \hspace{1.5cm} -\frac{2W_2(1+W_1) \, [{\rm Im}(\alpha-\beta )]^2 }{1+W_1+W_2+W_1W_2(1+16K^2)}
     \nonumber \\
&& \hspace{1.5cm} -\frac{16W_1W_2K \, {\rm Re}(\alpha -\beta) \, {\rm Im}(\alpha-\beta)} {1+W_1+W_2+W_1W_2(1+16K^2)} \bigg\}.
    \label{eq:Q-expand}
\end{eqnarray}
Comparing this formula with Eq.\ (\ref{eq:Q-gauss}) for the Gaussian state, we see that the formulas coincide if
\begin{align}
&D_0 = \frac{1}{8W_2} + \frac{W_1}{8} + 2K^2W_1,
    \label{D0-conv-app} \\
&b = \frac{1}{4} \sqrt{\Big( \frac{1}{2W_2}+\frac{W_1}{2}+8K^2W \Big)^2 -\frac{W_1}{W_2}},
     \label{b-conv-app}\\
&\theta_0 = \arctan\Big( \frac{8KW_1W_2}{1-W_1W_2+16K^2W_1W_2} \Big)
    \nonumber \\
& \hspace{0.5cm} + (\pi /2)\,[1-{\rm sign}(1-W_1W_2+16K^2W_1W_2)] ,
     \label{theta0-conv-app}
\end{align}
where we use notation $\theta_0$ instead of $\theta$ to remind that we consider the special case of a real positive $\beta$. Note that Eqs.\ (\ref{D0-conv-app})--(\ref{theta0-conv-app}) coincide with Eqs.\ (\ref{eq:conv-D})--(\ref{eq:conv-theta}) in the case of a real positive $\beta$.

For a complex $\beta$, it is also possible to use the second-order expansion of Eq.\ (\ref{eq:Q-full}); however, it is easier to use the fact that dependence of $Q(\alpha )$ on the complex phase $\phi_\beta$ in Eqs.\ (\ref{Q-FG-1-app})--(\ref{C-def}) comes only from the combination $\phi_\alpha-\phi_\beta$.
Therefore, the $Q$-function of the Fock-space Gaussian state does not change in the transformation $\beta \to |\beta|$, $\alpha \to e^{-i\phi_\beta}\alpha$, so for a complex $\beta$ we can still  use Eq.\ (\ref{eq:Q-expand}) with the substitution $(\alpha-\beta)\to e^{-i\phi_\beta}(\alpha-\beta)$.
Using this substitution in the equivalent Eq.\ (\ref{eq:Q-gauss}), we easily find that it results in replacing the angle $\theta_0$ (for real $\beta$) with
\begin{equation}\label{theta-theta0-app}
	\theta = \theta_0+2 \phi_\beta ,
\end{equation}
while the parameters $D_0$ and $b$ do not change.
Another way to obtain Eq.\ (\ref{theta-theta0-app}) is to note that the parameters $W_1$, $W_2$, and $K$ of the Fock-space Gaussian state do not change when the phase space is rotated (i.e., $\beta \to e^{i\Delta\phi} \beta$, $\alpha \to e^{i\Delta\phi} \alpha$), while for the Gaussian state this results in the change $\theta \to \theta +2\Delta\phi$ with unchanged parameters $D_0$ and $b$ (see Fig.\ \ref{fig:quad}).
Therefore, we can first rotate the phase space clockwise by the angle $\phi_\beta$ (to make $\beta$ real), then convert parameters $W_1$, $W_2$, and $K$, into $D_0$, $b$, and $\theta_0$ using Eqs.\ (\ref{D0-conv-app})--(\ref{theta0-conv-app}), and then move the phase space back by counterclockwise rotation with the same angle $\phi_\beta$, which results in $\theta$ change (\ref{theta-theta0-app}).

Thus we have derived the conversion relations (\ref{eq:conv-D})--(\ref{eq:conv-theta}) between the Gaussian and Fock-space Gaussian states ($\beta$ does not change).
Note that our derivation relied on the fact that the Husimi $Q$-function uniquely defines a quantum state \cite{Gerry2005}.
Since Eq.\ (\ref{eq:Q-expand}) is only an approximation, a Fock-space Gaussian state is not exactly equal to a Gaussian state.
However, the accuracy of the conversion improves at larger $|\beta|$, approaching exact equivalence in the limit $|\beta|\to \infty$.
Numerical results in Sec.\ \ref{subsec:conversion-fidelity} show that infidelity of the conversion scales as $|\beta|^{-2}$.

\section{Steady-state squeezing and heating} \label{app:steady}

In this appendix, we derive results for $D_0$, $b$, and $\theta$ in the steady state. The parameters $r$ and $n_{\rm th}$ can be then calculated using Eq.\ (\ref{eq:nth-r}). The squeezing factor is $[4(D_0-b)]^{-1}$, the effective temperature $T_{\rm eff}$ is given by $\coth (\omega_{\rm r0}/2T_{\rm eff})=4\sqrt{(D_0+b)(D_0-b)}$.  All variables discussed in this appendix are only for the steady state.

The steady-state value of $\beta$ can be calculated from Eq.\ (\ref{eq:hybrid-beta}); in general it does not have an analytical expression. Note that
    \begin{equation}
    \varepsilon/\beta = \omega_{\rm d}- \omega_{\rm r} (|\beta|^2) +i\kappa/2,
    \end{equation} so ${\rm Re}(\varepsilon/\beta)$ can be positive or negative, depending on detuning.

From Eqs.\ (\ref{eq:hybrid-D0}) and (\ref{eq:hybrid-b}) in the steady state we find
    \begin{align}
&    D_0 = \frac{\coth(\omega_{\rm r0}/2T_{\rm b})}{4} \,  \frac{1}{1-[2\eta_\beta |\beta|^2\sin (\Delta \theta)/\kappa ]^2 } ,
    \label{steady-D0}\\
& b= \frac{\coth(\omega_{\rm r0}/2T_{\rm b})}{4} \, \frac{2\eta_\beta |\beta|^2\sin (\Delta \theta)/\kappa}{1-[2\eta_\beta |\beta|^2\sin (\Delta \theta)/\kappa ]^2} ,
    \label{steady-b}\end{align}
where $\eta_\beta=d\omega_{\rm r} (n)/dn|_{n=|\beta|^2}$ is the steady-state nonlinearity. To obtain explicit analytics for $D_0$ and $b$, we still need to find $\sin (\Delta \theta )$. For that we can substitute the ratio $b/D_0=2\eta_\beta |\beta|^2\sin (\Delta \theta)/\kappa$ into Eq.\ (\ref{eq:hybrid-theta}) in the steady state, thus obtaining
    \begin{equation}
    \tan (\Delta\theta) = \frac{\kappa /2}{\eta_\beta |\beta|^2 -{\rm Re} (\varepsilon /\beta)}.
    \end{equation}
Since $\eta_{\beta}\sin (\Delta\theta)\geq 0$ (because $b\geq 0$), we can use
    \begin{equation}
    \sin (\Delta \theta) = {\rm sign} (\eta_\beta) \sqrt{\frac{(\kappa/2)^2}{(\kappa/2)^2+[\eta_\beta |\beta|^2 -{\rm Re} (\varepsilon /\beta)]^2}}  \quad
    \end{equation}
in Eqs.\ (\ref{steady-D0}) and (\ref{steady-b}).

The angle $\theta$ can be calculated as
    \begin{align}
& \theta =2\,{\rm arg} (\beta) + {\rm arctan} \left( \frac{\kappa /2}{\eta_\beta |\beta|^2 -{\rm Re} (\varepsilon /\beta)} \right)
    \nonumber \\
&\hspace{0.5cm} + (\pi/2) \{1-{\rm sign [|\beta|^2- \eta_\beta^{-1} {\rm Re} (\varepsilon /\beta)}]\} .
    \label{steady-theta}\end{align}

\vspace{0.3cm}

These results can be compared with results of  Ref.\ \cite{Drummond1980} in  the case of Kerr nonlinearity (Duffing oscillator), $H_{\rm r}^{\rm lf}= \omega_{\rm r0} a^\dagger a +(\eta /2) (a^\dagger)^2 a^2$, which is equivalent to our Hamiltonian when $\omega_{\rm r}=\omega_{\rm r0}+n\eta$ [see Eq.\ (\ref{Kerr})], so that $\eta_\beta =\eta={\rm const}$. In this case Eq.\ (4.4) of Ref.\ \cite{Drummond1980} (converted into our notations) gives
    \begin{align}
    & \langle a^2\rangle - \beta^2 = - \frac{\eta \beta^2 (\omega_{\rm r0}+2\eta |\beta|^2 -\omega_{\rm d} +i\kappa/2)(1+2n_{\rm b})}{2\lambda} ,
    \\
    & \langle a^\dagger a \rangle -|\beta|^2 =\frac{\eta^2 |\beta|^4 (1+2 n_{\rm b})}{2\lambda} +n_{\rm b} ,
    \\
    & \lambda = (\omega_{\rm r0}+2\eta |\beta|^2-\omega_{\rm d})^2 + \kappa^2/4 -\eta^2 |\beta|^4 .
    \end{align}
From these values, $D_0$, $b$, and $\theta$ can be obtained using Eqs.\ (\ref{eq:a-to-D})--(\ref{eq:a-to-theta}) [also, Eq.\ (\ref{nbar-via-D}) gives $\langle a^\dagger a \rangle -|\beta|^2=2D_0-1/2$].

We have numerically compared these results  with our Eqs.\ (\ref{steady-D0}), (\ref{steady-b}), and (\ref{steady-theta}) and found that they coincide for all parameters, which we checked. Thus, for the steady state in the case of Kerr nonlinearity, our results for squeezing and heating agree with results of Ref.\
\cite{Drummond1980} (note that the terminology of squeezing and/or  heating was not used in Ref.\ \cite{Drummond1980}).

Our steady-state results for a Duffing oscillator in the limit of small dissipation ($\kappa \to 0$) can also be directly compared with the analytical results presented in Secs.\ 2.1 and 2.5 of Ref.\ \cite{Dykman-2012}.
In this case the squeezing and heating are determined only by the parameter combination $\varepsilon^2\eta /(\omega_{\rm d}-\omega_{\rm r0})^3$ (which was called $\beta$ in Ref.\ \cite{Dykman-2012}). Results of Ref.\ \cite{Dykman-2012} show that the squeezing parameter $\xi=r e^{i\theta}$ is real and equals
    \begin{equation}
\xi=\frac{1}{4} \ln\frac{3Q^2 -1}{Q^2-1},
    \label{xi-app}\end{equation}
where $Q$ satisfies equation
    \begin{equation}
Q(Q^2-1)=\sqrt{\varepsilon^2\eta /(\omega_{\rm d}-\omega_{\rm r0})^3}.
    \end{equation}
Here in the case $\varepsilon^2\eta /(\omega_{\rm d}-\omega_{\rm r0})^3>4/27$, there is only one real solution for $Q$. The range $0<\varepsilon^2\eta /(\omega_{\rm d}-\omega_{\rm r0})^3<4/27$  corresponds to bistability, and there are three real solutions for $Q$, with the largest value corresponding to the upper bistability branch and the middle value for the lower branch. In the case  $\varepsilon^2\eta /(\omega_{\rm d}-\omega_{\rm r0})^3<0$, we need to use the purely imaginary solution for $Q$.

The angle $\theta$ in this limit is zero (squeezing is in phase with the drive), except $\theta=\pi$ for the lower bistability branch  (then $\xi<0$). The number of thermal photons is \cite{Dykman-2012}
     \begin{equation}
     n_{\rm th}=n_{\rm b}+(2n_{\rm b}+1)\sinh ^2 r.
    \label{n-th-app}\end{equation}
We have numerically compared Eqs.\ (\ref{xi-app}) and (\ref{n-th-app}) for $\xi$ and $n_{\rm th}$ with our results following from Eqs.\ (\ref{steady-D0}), (\ref{steady-b}), and (\ref{steady-theta}). As expected, we have found that they coincide in the limit $\kappa \to 0$ for a fixed value of $\varepsilon^2\eta /(\omega_{\rm d}-\omega_{\rm r0})^3$. Thus, our results agree with the results of Ref.\ \cite{Dykman-2012}.



%

\end{document}